\documentclass[acmsmall, nonacm
]{acmart}

\AtBeginDocument{%
  }

\setcopyright{acmlicensed}
\copyrightyear{2018}
\acmYear{2018}
\acmDOI{XXXXXXX.XXXXXXX}

\acmConference[Conference acronym 'XX]{Make sure to enter the correct
  conference title from your rights confirmation emai}{June 03--05,
  2018}{Woodstock, NY}
\acmISBN{978-1-4503-XXXX-X/18/06}

\usepackage{listings}
\usepackage{subcaption}
\usepackage{graphicx}
\usepackage{wrapfig}
\usepackage{tabularx}
\usepackage{multirow}
\usepackage{makecell}
\usepackage{xspace}
\usepackage[ruled,vlined]{algorithm2e}
\usepackage{booktabs}
\usepackage{tabularx}
\usepackage{hyperref}
\usepackage{enumitem}
\usepackage{xcolor}

\lstdefinelanguage{json}{
    basicstyle=\normalfont\ttfamily,
    showstringspaces=false,
    breaklines=true,
    frame=single,
    string=[b]",
    morestring=[b]',
    literate=
     *{0}{{{\color{blue}0}}}{1}
      {1}{{{\color{blue}1}}}{1}
      {2}{{{\color{blue}2}}}{1}
      {3}{{{\color{blue}3}}}{1}
      {4}{{{\color{blue}4}}}{1}
      {5}{{{\color{blue}5}}}{1}
      {6}{{{\color{blue}6}}}{1}
      {7}{{{\color{blue}7}}}{1}
      {8}{{{\color{blue}8}}}{1}
      {9}{{{\color{blue}9}}}{1}
      {:}{{{\color{red}:}}}{1}
      {,}{{{\color{red},}}}{1}
}

\usepackage[most]{tcolorbox}
\tcbset{
  mybox/.style={
    colback=gray!10,
    colframe=black!80!white,
    fonttitle=\bfseries,
    coltitle=black,
    boxrule=0.4mm,
    width=\textwidth,
    arc=0.5mm,
    boxsep=0.5pt,
    left=0.5pt,
    right=0.5pt,
    fontupper=\bfseries
  }
}

\begin{document}


\title{LLM-based Unit Test Generation via Property Retrieval}

\author{Zhe Zhang}
\affiliation{%
 \institution{Beihang University}
 \orcid{0009-0007-4089-4087}
 \country{China}}
 \email{zhangzhe2023@buaa.edu.cn}

\author{Xingyu Liu}
\affiliation{%
 \institution{Beihang University}
 \country{China}}
 \email{lxingyu@buaa.edu.cn}

\author{Yuanzhang Lin}
\affiliation{
 \institution{Beihang University}
 \orcid{0009-0009-6294-326X}
 \country{China}}
 \email{yuanzhanglin@buaa.edu.cn}

\author{Xiang Gao}
\authornote{Corresponding author.}
\affiliation{
 \institution{Beihang University}
 \orcid{0000-0001-9895-4600}
 \country{China}}
\email{xiang_gao@buaa.edu.cn}

\author{Hailong Sun}
\affiliation{
 \institution{Beihang University}
 \orcid{0000-0001-7654-5574}
 \country{China}}
\email{sunhl@buaa.edu.cn}

\author{Yuan Yuan}
\affiliation{
 \institution{Beihang University}
 \country{China}}
\email{yuan21@buaa.edu.cn}







\renewcommand{\shortauthors}{Zhang et al.}
\newcommand{\delete}[1]{\textcolor{red}{#1}}

\newcommand{\nbc}[3]{
	{\colorbox{#3}{\bfseries\sffamily\scriptsize\textcolor{white}{#1}}}
	{\textcolor{#3}{\sf\small$\blacktriangleright$\textit{#2}$\blacktriangleleft$}}
}

\newcommand{\xiang}[1]{\nbc{GX}{#1}{cyan}}
\newcommand{\zhangzhe}[1]{\nbc{ZZ}{#1}{blue}}
\newcommand{\zxt}[1]{\nbc{XT}{#1}{magenta}}

\newcommand{\toolname}[0]{{APT}\xspace}

\renewcommand{\footnoterule}{
  \kern -3pt
  \hrule width 0.4\columnwidth height 0.4pt
  \kern 2.6pt
}
\renewcommand{\arraystretch}{1.2}

\newcommand{\add}[1]{\textcolor{}{#1}}
\begin{abstract}

Automated unit test generation has been widely studied, with Large Language Models (LLMs) recently showing significant potential. LLMs like GPT-4, trained on vast text and code data, excel in various code-related tasks, including unit test generation. However, existing LLM-based tools often narrow their focus to general code context, overlooking broader task-specific context, such as leveraging existing tests in unit test generation tasks. Moreover, in the context of unit test generation, these tools prioritize high code coverage, often at the expense of practical usability, correctness, and maintainability.

In response, we propose \textit{Property-Based Retrieval Augmentation}, a novel mechanism that extends LLM-based Retrieval-Augmented Generation (RAG) beyond basic vector, text similarity, and graph-based methods. Our approach considers task-specific context in the code domain and introduces a tailored property retrieval mechanism that leverages the unique characteristics of code. Specifically, in the unit test generation task, we account for the unique structure of unit tests by dividing the test generation process into \textit{Given}, \textit{When}, and \textit{Then} phases. When generating tests for a focal method, we not only retrieve general context for the code under test but also consider task-specific context such as pre-existing tests of other methods, which can provide valuable insights for any of the \textit{Given}, \textit{When}, and \textit{Then} phases. This forms \textit{property relationships} between focal method and other methods, thereby expanding the scope of retrieval beyond traditional RAG. 
We implement this approach in a tool called \emph{\toolname}, which sequentially performs preprocessing, property retrieval, and unit test generation, using an iterative strategy where newly generated tests guide the creation of subsequent ones. We evaluated \toolname on 12 open-source projects with 1515 methods, and the results demonstrate that \toolname consistently outperforms existing tools in terms of correctness, completeness, and maintainability of the generated tests. Moreover, we introduce a novel code-context-aware retrieval mechanism for LLMs beyond general context, offering valuable insights and potential applications for other code-related tasks.
\end{abstract}

\maketitle

\section{Introduction}

Automated unit test generation has been extensively studied, resulting in a variety of approaches. Prominent methods include random-based strategies \cite{pacheco2007randoop, davis2023nanofuzz, wei2022free} and constraint-driven techniques, such as DART \cite{godefroid2005dart} and KLEE \cite{ma2015grt}. Search-Based Software Testing (SBST) techniques \cite{chen2023compiler, feldmeier2022neuroevolution, lemieux2023codamosa, lin2023route, sun2023evolutionary, yandrapally2022fragment, zhou2022selectively, lukasczyk2022pynguin}, exemplified by tools like EvoSuite \cite{fraser2011evosuite}, aim to improve the effectiveness of test generation but still struggle with large search spaces and high computational costs \cite{mcminn2011search}. Recently, deep learning-based approaches, such as AthenaTest \cite{tufano2020unit}, have emerged, utilizing neural models to generate more diverse test inputs and better capture the functional intent of code \cite{blasi2022call, feldmeier2022neuroevolution, wang2020automatic, ye2023generative, zhao2022avgust}.

Additionally, Large Language Models (LLM)-based techniques are becoming increasingly popular in automated testing \cite{guilherme2023initial, deng2023large, lemieux2023codamosa}. Tools like ChatUniTest \cite{xie2023chatunitest} and ChatTester \cite{yuan2023no} leverage ChatGPT for unit test generation.
Relying on proper prompt engineering and rich code context, LLM-based tools often outperform traditional methods like SBST in certain cases. 
Hybrid approaches~\cite{yang2024enhancing, wang2024hits} combine the strengths of LLMs and traditional techniques to improve test quality.
For instance, TELPA \cite{yang2024enhancing}, uses program analysis with LLMs by integrating refined counterexamples into prompts, guiding the LLM to generate diverse tests for hard-to-cover branches, 
and HITS \cite{wang2024hits} decomposes focal methods into slices and asks the LLM to generate tests slice by slice, improving coverage for complex methods.



\paragraph{\textbf{Observation 1: Focus on Code Itself in LLM-based Code Tasks, Overlooking Task-Specific Context}}

Existing LLM-based code generation tasks, particularly Retrieval-Augmented Code Generation (RACG), enhance the capabilities of LLMs by integrating relevant code snippets or structures from code repositories. However, these approaches often concentrate solely on contextualizing the original code using various retrieval techniques, such as similarity-based retrieval (e.g., BM25 \cite{jimenez2023swe}), AST-based retrieval \cite{zhang2024autocoderover}, graph-based retrieval \cite{liu2024codexgraph}, iterative file-based retrieval \cite{zhang2023repocoder}, and vector similarity-based retrieval \cite{pan2024enhancing}. Despite their effectiveness, these methods do not fully account for the specific context of the code task at hand.

In the case of unit test generation, when generating tests for a focal method, can we go beyond just the general code context?
The broader context, including pre-existing test cases, can greatly improve the quality and relevance of the generated tests. This underscores the necessity for developing retrieval mechanisms that are specifically designed to support the nuances of different code tasks, thereby enriching the overall process of automated test generation.

{\textit{\textbf{Key question:}}
{How can we develop a retrieval mechanism that leverages task-specific information, such as considering existing test cases in the context of unit test generation, to enhance LLM-based code generation for specific tasks?}

\paragraph{\textbf{Observation 2: Practical Problem in Unit Test Generation}}

Despite advancements in automated test generation, many tools remain focused on maximizing coverage metrics. However, a recent study by Yu et al. \cite{yu2024practitioners} reveals that practitioners value other attributes more.
Although coverage is often emphasized in academic discussions, the study highlights that the correct rate (the proportion of generated test code that is syntactically correct, compilable, and runnable) and passing rate (the proportion of test cases that accurately reflect
the requirements)
are the most critical factors for adopting test generation tools in the real world. 
Moreover, the understandability (an aspect of auto-generated code that allows developers to quickly read and understand it) \cite{grano2018empirical} and maintainability (developers can easily modify, extend, and adapt generated test cases over time) of the generated tests are also highly prioritized, reflecting the practical needs of developers in real-world settings.
Furthermore, their study reveals that a majority of test generation tools (70\%) focus predominantly on the code under test. 
For instance, ChatUniTest~\cite{xie2023chatunitest} incorporates an adaptive focal context mechanism that captures useful context within prompts. 
However, this often does not fully align with developers' expectations, as these tools frequently overlook broader aspects of the codebase, such as method relationships and existing test cases.

{\textit{\textbf{Key question:}}
{How can we develop an approach that not only maximizes coverage but also produces unit tests that are correct, highly understandable, and maintainable?}

\paragraph{\textbf{Our Approach}}
In contrast to conventional approaches that focus solely on the code under test, our method incorporates a broader perspective by considering both the static context of the code under test (e.g., variables, function references) and \textbf{existing test cases} in the repository. 
We extend the concept of context retrieval for LLMs by defining \textit{property relationships}. Specifically, unit tests are usually structured into three key phases: \textit{Given}, \textit{When}, and \textit{Then} \cite{GWT}. For each phase, we search for methods that can provide valuable insights during test generation, based on the characteristics of the corresponding phase.

Let us consider a simple example of a property relationship. Suppose we are generating a unit test for \texttt{decode} method (Base64 decoding), and there already exists a test case for \texttt{encode} method (Base64 encoding). Although \texttt{encode} and \texttt{decode} are inverse operations, the test case of \texttt{encode} can still provide useful references at each phase of the test generation of \texttt{decode}:


\begin{itemize}[leftmargin=*]
    \item \textbf{Given Phase}: The \texttt{encode} test case provides object initialization and input setup, which can give reference for \texttt{decode}, as both methods process similar data structures (e.g., handling strings and converting them to/from Base64 format).
    
    \item \textbf{When Phase}: Although \texttt{encode} and \texttt{decode} perform inverse actions, the method invocation in \texttt{encode}, including how input data is handled, offers valuable insights for calling \texttt{decode}, particularly in the way the data is prepared before processing.
    
    \item \textbf{Then Phase}: The assertions and exception handling in \texttt{encode} guide the verification process for \texttt{decode}, 
    such as output validation’s format, boundary checking (e.g., verifying the behavior when the input is an empty), and exception handling (e.g., reusing the same logic for catching and handling invalid Base64 input).
\end{itemize}

Our proposed method systematically identifies such references during test generation for a focal method by looking for task-specific contexts like these. By abstracting this process through the definition of \textit{property relationships}, our approach enables the retrieval of task-specific contexts, enhancing the effectiveness of test case generation.

Based on our approach, we propose \toolname, which preprocesses the project by parsing the code and converting it into a relational schema stored in a \textit{Metainfo Database}. This supports both retrieval and test generation. Next, \toolname analyzes existing test cases, extracts \textit{test bundles}—the core reference units for generating new tests—and maps them to focal methods. Additionally, \toolname introduces a \textit{Static Context Retrieval} mechanism to streamline the retrieval and generation processes.
Then, \toolname executes two main steps:


\begin{itemize}[leftmargin=*]
    \item \textbf{Retrieval}. After preprocessing, \toolname performs property retrieval. For a given focal method, it uses \textit{property analysis} to find methods with property relationships within the class. It then extends to class inheritance and interface implementation to perform \textit{property deduction}, expanding the search to the entire repository. The relevant methods and their \textit{test bundles} are retrieved.
    
    \item \textbf{Generation}. \toolname ranks the retrieved methods and \textit{test bundles} by relevance and confidence, combining them with the static context of the focal method. The most relevant components are then provided to the LLM for test case generation. Additionally, \toolname applies an \textit{Iterative Strategy}, where newly generated test cases guide the generation of subsequent tests, maximizing the number of correct test cases.
\end{itemize}

In Summary, we propose a novel approach, \textit{Property-Based Retrieval Augmentation for Unit Test Generation}.
This method defines \textit{property relationships} between methods, 
\toolname extends traditional vector-similarity-based retrieval by incorporating code-specific relationships such as behavioral similarities, structural similarities, dependencies, and inheritance and interface implementations among classes. By leveraging these property relationships and contextual information, our approach significantly improves the \textbf{correctness}, \textbf{completeness}, and \textbf{maintainability} of the generated unit tests. This paper offers several key contributions:
\begin{itemize}[leftmargin=*]
    \item We are the first to propose a \textit{property-based retrieval} approach for unit test generation. By decomposing the test generation process into the \textit{GWT} phases and defining properties between methods and property retrieval, we leverage existing test cases to generate unit tests for focal methods. This extends the traditional Retrieval-Augmented Generation (RAG) paradigm to incorporate code-specific property features.
    
    \item We propose \toolname, which integrates preprocessing, retrieval, and generation to produce high-quality unit tests. Additionally, our iterative strategy maximizes the number of focal methods that can be covered by leveraging existing test cases with property relationships.
    
    \item We conduct comprehensive evaluations of \toolname on 12 open-source projects with 1515 methods. The results show that \toolname consistently outperforms existing methods in terms of \textbf{correctness}, \textbf{completeness}, and \textbf{maintainability} of the generated tests.
\end{itemize}

\section{Motivating Example}
In this section, we first introduce an example of \textit{property relationship} that exists at the Class-Level. We then extend this relationship to the Repo-Level.

\subsection{Class-Level Enhancement}

To illustrate motivations behind our approach, we take as an example the \texttt{reset} method from the \texttt{RedissonLongAdder} class in the \texttt{Redisson} repository \citep{Redisson}, a widely used Java client for Redis with over 23k stars on GitHub. 
We start by generating unit tests for the \texttt{reset} method.
\begin{figure}[htbp]
    \centering
    \vspace{-6pt}
    \begin{lstlisting}[language=Java, basicstyle=\tiny, numbers=left, breaklines=true, postbreak=\mbox{\textcolor{red}{$\hookrightarrow$}\space}]
public RFuture<Void> reset() {
    String id = getServiceManager().generateId();
    RSemaphore semaphore = getSemaphore(id);
    RFuture<Long> future = topic.publishAsync(CLEAR_MSG + ":" + id);
    CompletionStage<Void> f = future.thenCompose(r -> semaphore.acquireAsync(r.intValue()))
                                    .thenCompose(r -> semaphore.deleteAsync().thenApply(res -> null));
    return new CompletableFutureWrapper<>(f);
}
    \end{lstlisting}
    \vspace{-6pt}
    \caption{The method to be tested from the \texttt{RedissonLongAdder} class: \texttt{reset} method}
    \label{fig:reset_method}
    \vspace{-5pt}
\end{figure}


The \texttt{reset} method is intended to set a counter to zero in a distributed system, where multiple instances might share the same counter. This operation guarantees that the reset is applied to all instances.
It begins by generating a unique identifier (line 2) to fetch a \texttt{semaphore}, which is a synchronization tool (line 3). The method then sends a clear message to a \texttt{topic}, a communication channel, which returns a \texttt{future} representing the asynchronous operation (line 5). Upon successful message delivery, it acquires and deletes the semaphore (lines 6-7), ensuring a controlled reset. The entire process is encapsulated in a CompletableFutureWrapper, facilitating safe and efficient asynchronous handling, crucial for managing concurrency in distributed systems.

When using existing unit test generation tools~\cite{xie2023chatunitest, yuan2023no}, they strive to search for the general context (e.g., the definition of variables and the invoked methods that are defined outside the\begin{wrapfigure}{r}{0.55\textwidth}
    \centering
    \vspace{-15pt}
    \begin{lstlisting}[language=Java, basicstyle=\tiny, numbers=left, breaklines=true, numbersep=2pt, postbreak=\mbox{\textcolor{red}{$\hookrightarrow$}\space}]
@BeforeEach
public void setUp() {
    redissonMock = mock(RedissonClient.class);
    redissonLongAdder = new RedissonLongAdder(mock(CommandAsyncExecutor.class), "test", redissonMock);
}
@Test
public void testReset() {
    RSemaphore semaphoreMock = mock(RSemaphore.class);
    ......
    RFuture<Void> result = redissonLongAdder.reset();

    verify(semaphoreMock).acquireAsync(anyInt());
    verify(semaphoreMock).deleteAsync();
    Assert.assertThat(result).isNotNull();
}
    \end{lstlisting}
    \vspace{-6pt}
    \caption{The Unit Test for \texttt{reset} generated by GitHub Copilot}
    \vspace{-5pt}
    \label{fig:reset_method2}
\end{wrapfigure} method) 
of the code under test in order to better understand what the method does, thus helping to generate test cases. 
For example, When generating tests using GitHub Copilott~\citep{GithubCopiot}, it looks up the source code of functions \texttt{getServiceManager}, \texttt{generateId}, and \texttt{getSemaphore}, determining where \texttt{topic} is defined and its type to have a better understanding of the method being tested.
As shown in Figure \ref{fig:reset_method2}, the test generated by GitHub Copilot first sets up the environment (lines 1-5) by mocking \texttt{RedissonClient} and creating \texttt{RedissonLongAdder} with mocked dependencies.
Then, the test directly calls \texttt{reset()} (lines 11) and verifies that the result of \texttt{reset()} is not null (lines 15).
Moreover, the test also verifies that \texttt{acquireAsync} and \texttt{deleteAsync} can be correctly invoked.
The test primarily focuses on mocking external objects and verifying their interactions.


However, when compiling this test case, it failed due to the heavy reliance on mocked external services.  
Additionally, the test fails to validate the actual core functionality of the \texttt{reset()} method. The core functionality of \texttt{reset()} is to reset the internal state of the \texttt{LongAdder}, ensuring that any accumulated values are reset to zero. However, the generated test is focused primarily on confirming that external methods were invoked, without verifying whether the internal state reset was correctly applied.

This raises an important question: \textit{Is understanding the \texttt{reset()} method and looking up relevant external references sufficient to construct meaningful and effective unit tests}? 
In this case, the test lacks a deeper understanding of the method's internal logic and only focuses on superficial features, which leads to incomplete and ineffective test coverage for the method's core functionality.

\textbf{Our Insight}. 
Our tool \toolname focuses on identifying methods that have a \textit{property relationship} with the focal method. 
If these related methods already have test cases, they can provide valuable guidance and reference during the test generation process for the focal method.
Specifically, when generating unit tests for the focal method \texttt{reset}, 
through \textit{property analysis}, \toolname identifies  a \textit{property relationship} between \texttt{reset} method and \texttt{sum} method from class \texttt{RedissonLongAdder}.
Although their functionalities are semantically different, \toolname observes significant similarities in the GWT stages of their unit tests.
The test cases of both methods are shown in Figure~\ref{fig:testSum_and_testReset}, their similarities are summarized as follows:
\begin{enumerate}[leftmargin=*]
    \item \textbf{Given}: Both \texttt{reset} and \texttt{sum} require initializing the \texttt{RedissonLongAdder} class and configuring it appropriately before performing any operations.
    \item \textbf{When}: For \texttt{sum}, the method is invoked to calculate the cumulative sum of values added to the adder, while \texttt{reset} is invoked to reset the value back to zero. Despite these differing outcomes, the structure of invoking the method follows a similar pattern in both cases.
    \item \textbf{Then}: In both cases, the assertions verify the final state of the adder. For \texttt{sum}, the assertion checks that the value returned matches the expected sum, whereas for \texttt{reset}, it should ensure that the value is reset to zero, confirming the correct functionality of both methods.
\end{enumerate}

\begin{figure}[h]
    \vspace{-6pt}
    \centering
    \begin{subfigure}[b]{0.48\textwidth}
        \begin{lstlisting}[language=Java, basicstyle=\scriptsize, breaklines=true, escapeinside={||}, postbreak=\mbox{\textcolor{red}{$\hookrightarrow$}\space}]
@Test
public void testSum() {
    var adder1 = redisson.getLongAdder("test1");
    var adder2 = redisson.getLongAdder("test1");
    
    adder1.add(2);
    adder2.add(4);
    
    assertThat(adder1.|\textbf{sum}()|).isEqualTo(6);
    assertThat(adder2.|\textbf{sum}()|).isEqualTo(6);
    adder1.destroy();
    adder2.destroy();
}
        \end{lstlisting}
        \caption{The Unit Test of \texttt{sum}}
        \label{fig:testSum}
    \end{subfigure}
    \hspace{0.02\textwidth} 
    \begin{subfigure}[b]{0.48\textwidth}
        \begin{lstlisting}[language=Java, basicstyle=\scriptsize, breaklines=true, escapeinside={||}, postbreak=\mbox{\textcolor{red}{$\hookrightarrow$}\space}]
@Test
public void testReset() {
    var adder1 = redisson.getLongAdder("test1");
    var adder2 = redisson.getLongAdder("test1");
    adder1.add(2);
    adder2.add(4);
    adder1.|\textbf{reset}()|;
    
    assertThat(adder1.sum()).isZero();
    assertThat(adder2.sum()).isZero();
    adder1.destroy();
    adder2.destroy();
}
        \end{lstlisting}
        \caption{The Unit Test of \texttt{reset}}
        \label{fig:testReset}
    \end{subfigure}
    \vspace{-6pt}
    \caption{Two examples of Redisson LongAdder methods demonstrating the reuse of patterns from the \texttt{sum} method's unit test to generate a unit test for the \texttt{reset} method. The figure illustrates the similarities in the initialization, invocation, and assertion phases of the \texttt{sum} and \texttt{reset} methods.}
    \vspace{-6pt}
    \label{fig:testSum_and_testReset}
\end{figure}

When \toolname identifies the \texttt{sum} method and observes that its unit test already exists, it uses \texttt{sum}'s tests as a reference to generate tests for the \texttt{reset} method. 
By leveraging the structure of \texttt{sum}'s test, \toolname simplifies and enhances the generation process for \texttt{reset}, reusing patterns from \texttt{sum}.
This allows the creation of a unit test (as shown in Figure~\ref{fig:testReset}) with similar initialization, invocation, and assertion logic, significantly reducing the complexity of test generation.




\subsection{Repo-Level Enhancement}
Further, in object-oriented programming, when we broaden our focus from an individual class to \begin{wrapfigure}{r}{0.45\textwidth}
    \centering
    \vspace{-15pt}
    \includegraphics[width=0.5\textwidth]{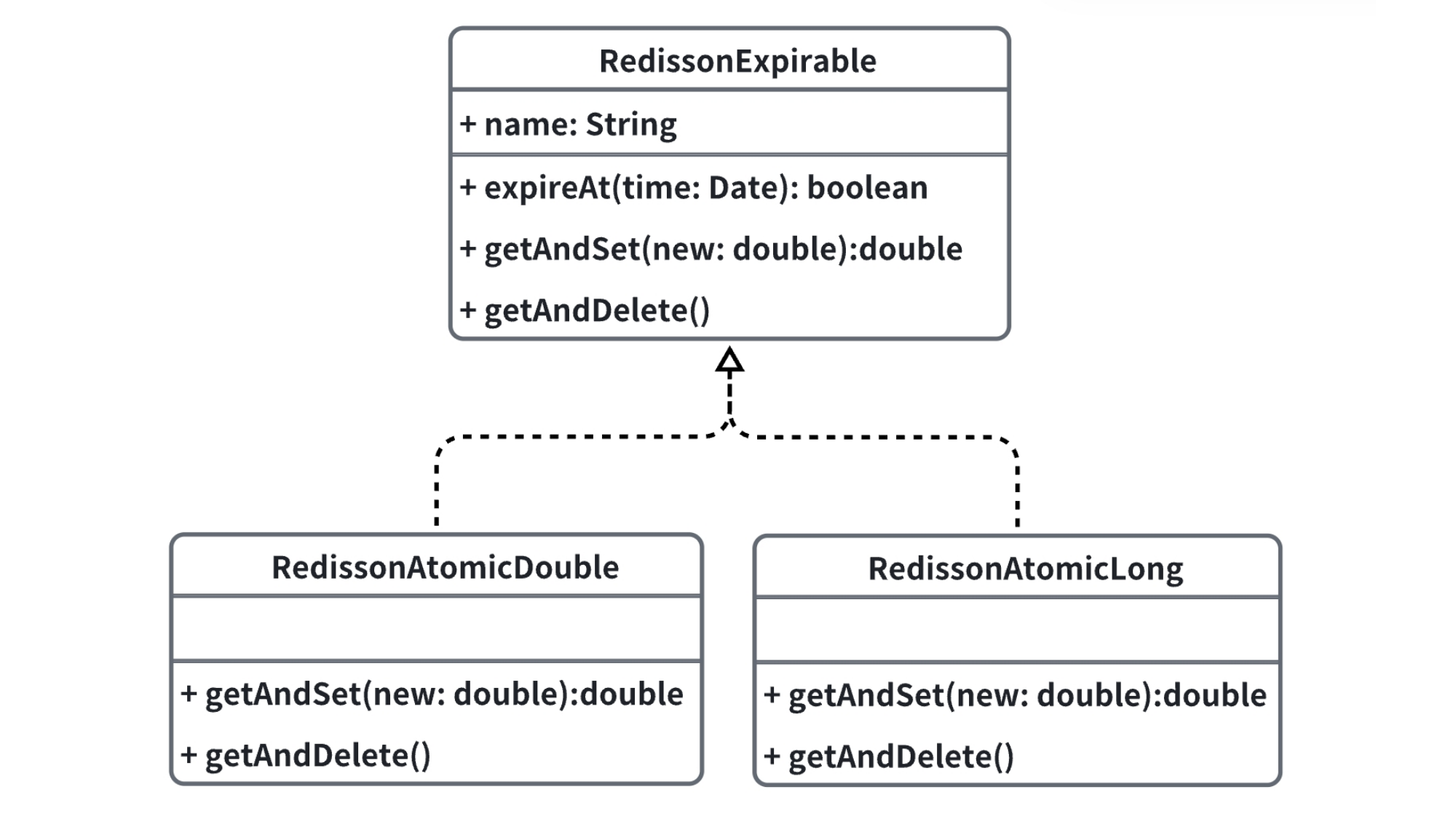} 
    \vspace{-8pt}
    \caption{RedissonAtomicDouble and RedissonAtomicLong both implementing the abstract class RedissonExpirable} 
    \vspace{-5pt}
    \label{fig:inherit}
\end{wrapfigure}the entire repository, we can identify more extensive attribute enhancement relationships:

\begin{itemize}[leftmargin=*]
    \item \textbf{Inheritance}: In cases of both concrete and abstract inheritance, methods between subclasses, parent classes, and sibling classes can be referenced when generating unit tests. 
    The relationships between different classes across the inheritance hierarchy can also be leveraged to enhance the unit test generation process. 
    
    \item \textbf{Common Interface Implementations}: 
    When multiple classes implement the same interface, the test generation process for these classes can benefit from mutual enhancement. For instance, the shared abstract behavior of two classes implementing a common interface can guide the generation of consistent and comprehensive unit tests. 
    
\end{itemize}

To demonstrate the Repo-Level test enhancement relationships, Figure \ref{fig:inherit} shows an example, where both class \texttt{RedissonAtomicLong} and \texttt{RedissonAtomicDouble} inherit from the base class \texttt{RedissonExpirable}.
The pre-existing test \texttt{testGetAndSet}, shown in Figure \ref{fig:RedissonAtomicDoubleTest}, from class \texttt{RedissonAtomicDouble} not only provides guidance for generating tests for other methods in the same class, such as \texttt{getAndDelete}, but also extends to its sibling class, \texttt{RedissonAtomicLong}.  
Specifically, in Figure ~\ref{fig:RedissonAtomicLongTest}, the \texttt{testGetAndSet} from \texttt{RedissonAtomicDoubleTest} helps generate unit tests for the \texttt{getAndSet} method in \texttt{RedissonAtomicLong} and can also assist in generating tests for \texttt{getAndDelete} in \texttt{RedissonAtomicLong}.


\begin{figure}[!thb]
  \centering
  \vspace{-6pt}
  \begin{subfigure}[b]{0.49\textwidth}
    \begin{lstlisting}[language=Java, basicstyle=\scriptsize, breaklines=true, escapeinside={||}, postbreak=\mbox{\textcolor{red}{$\hookrightarrow$}\space}]
public class RedissonAtomicDoubleTest{
  @Test
  public void testGetAndSet() {
    RAtomicDouble al = redisson.getAtomicDouble("test");
    assertThat(al.getAndSet(12)).isEqualTo(0);
  }
  @Test
  public void testGetAndDelete() {
    RAtomicDouble al = redisson.getAtomicDouble("test");
    al.set(1);
    assertThat(al.getAndDelete()).isEqualTo(1);
    assertThat(al.isExists()).isFalse();
    RAtomicDouble ad2 = redisson.getAtomicDouble("test2");
    assertThat(ad2.getAndDelete()).isZero();
  }
}
    \end{lstlisting}
    \caption{The Unit Test of \texttt{RedissonAtomicDouble}}
    \label{fig:RedissonAtomicDoubleTest}
  \end{subfigure}
  \hspace{0.01\textwidth} 
  \begin{subfigure}[b]{0.48\textwidth}
    \begin{lstlisting}[language=Java, basicstyle=\scriptsize, breaklines=true, escapeinside={||}, postbreak=\mbox{\textcolor{red}{$\hookrightarrow$}\space}]
public class RedissonAtomicLongTest{
  @Test
  public void testGetAndSet() {
    RAtomicLong al = redisson.getAtomicLong("test");
    Assert.assertEquals(0, al.getAndSet(12));
  }
  @Test
  public void testGetAndDelete() {
    RAtomicLong al = redisson.getAtomicLong("test");
    al.set(10);
    assertThat(al.getAndDelete()).isEqualTo(10);
    assertThat(al.isExists()).isFalse();
    RAtomicLong ad2 = redisson.getAtomicLong("test2");
    assertThat(ad2.getAndDelete()).isZero();
  }
}
    \end{lstlisting}
    \caption{The Unit Test of \texttt{RedissonAtomicLong}}
    \label{fig:RedissonAtomicLongTest}
  \end{subfigure}
  \caption{Unit tests for the \texttt{getAndSet} and \texttt{getAndDelete} methods in \texttt{RedissonAtomicDouble} and \texttt{RedissonAtomicLong}. The tests in \texttt{RedissonAtomicDoubleTest} can provide guidance for generating tests in \texttt{RedissonAtomicLongTest} due to the structural and behavioral similarities of the methods.}
  \label{fig:Repo_Level}
\end{figure}

Similarly, parent-child class relationships also exhibit similar test enhancement effects, where shared methods between parent and child classes benefit from the same mutual enhancement of unit tests. And when two classes implement a common interface, methods shared by these classes can follow the same pattern of test enhancement as sibling classes. In this case, tests generated for methods in one class can serve as valuable references for generating unit tests for the corresponding methods in the other implementing class, just like the relationship between sibling classes.

In summary, these property-based enhancement relationships extend across 
Class-Level and Repo-Level contexts. 
By taking these relationships into account, we can generate more accurate and effective unit tests for the focal method.




\section{PROBLEM FORMULATION}

In this section, we formally define the problem of generating test cases for focal methods in a code repository by leveraging existing test cases through property relationships.

\subsection{Formal Problem Definition}


Let \( Repo \) be a code repository containing a set of classes \( \mathit{C} = \{ c_1, c_2, \dots, c_k \} \). 
Methods in $Repo$ are denoted as \( M = \bigcup_{c \in \mathit{C}} M(c) \), where \( M(c) = \{ m_1, m_2, \dots, m_p \} \) is the set of methods in class \( c \).
Some methods have associated test cases, while others do not.
A set of existing test cases are donated as \( \mathit{TC} = \{ \mathit{tc}_1, \mathit{tc}_2, \dots \} \), each associated with some methods in \( M \).
Given a focal method \( m_t \in M \), 
our goal is to generate unit test(s) \( \mathit{UT}_{t} \) for \( m_t \) by:

\begin{enumerate}[leftmargin=*]
    \item \textbf{Property Definition:} Defining a \emph{property relation} \( \mathit{P} \subseteq M \times M \), where \( (m_t, m_r) \in \mathit{P} \) signifies that the method \( m_r \in M \) is related to \( m_t \) for the purpose of test case generation.
    \item \textbf{Property Retrieval:} Define the operation \( \mathit{PropertyRetrieval}(m_t) \), which takes a focal method \( m_t \) as input and retrieves the set of related methods \( \mathit{M}_r \) such that \( (m_t, m_r) \in \mathit{P} \). The result of this operation is denoted as \( \mathit{R}(m_t) \), representing the \emph{property relation method set}. This set contains all methods that have a property relationship with \( m_t \), along with the specific description of the property relationships (e.g., \textit{Given}, \textit{When}, \textit{Then} phases).

    \item \textbf{Generation:} 
    Utilize the property relation method set \( \mathit{R}(m_t) \) and the existing test cases \( \mathit{TC}_r \) associated with methods in \( \mathit{R}(m_t) \) to generate effective unit tests \( \mathit{TC}_t \) for the focal method \( m_t \).

\end{enumerate}

\subsection{Challenges}

Several challenges must be addressed to solve this problem effectively:

\subsubsection{Identifying Complex and Broader Property Relationships}
\label{sec:property_retrieval_challenge}

\begin{itemize}[leftmargin=*]
    \item \textbf{Complexity of Relationships}: Relationships across the \emph{Given}, \emph{When}, and \emph{Then} phases vary and are often complex, involving different types of connections such as structural, behavioral, and functional.

    \item \textbf{Broad Scope of Relationships}: These relationships can exist both within a class (intra-class) and across multiple classes (inter-class), making it crucial to consider relationships throughout the entire code repository.


    \item \textbf{Proximity and Relevance}: Not all relationships are equally useful. Some are more immediate, while others are more distant or indirect. Identifying the most relevant relationships that will contribute to effective test case generation is essential.

\end{itemize}

\subsubsection{Generating Effective Unit Tests}
\label{sec:ut_generation_challenge}

\begin{itemize}[leftmargin=*]

    \item \textbf{Capturing Contextual Dependencies from Existing Test Cases}: Test cases are not standalone but rely on specific setups (e.g., class members, fixtures, or external modules). Capturing these dependencies as part of the context is essential for generating correct and effective tests.

    \item \textbf{Integrating Property-Related Methods and Test Cases}: Once the relevant methods in \( \mathit{R}(m_t) \) are retrieved, the challenge lies in determining the priority of these methods and their associated test cases. This involves how to rank the methods based on their relevance and how to better integrate the existing test cases with the retrieved ones to provide the most effective context for the LLM during test case generation.

    
    \item \textbf{Maximizing Test Case Generation}: The goal is to generate unit tests for as many methods as possible by leveraging property relationships throughout the codebase. The challenge is to ensure that the \(\mathit{R}(m_t)\) relationships are effectively utilized not only for \(m_t\) but also to benefit as many methods as possible, allowing broader application of \emph{Property-Based Retrieval Augmentation for Unit Test Generation} across the entire project.

\end{itemize}

\section{METHODOLOGY}
We propose \toolname to address the problems and challenges previously outlined. The overall workflow of \toolname is illustrated in Figure \ref{fig:workflow}. 

\begin{figure}[t]
    \centering
    \vspace{-6pt}
    \includegraphics[width=0.9\textwidth]{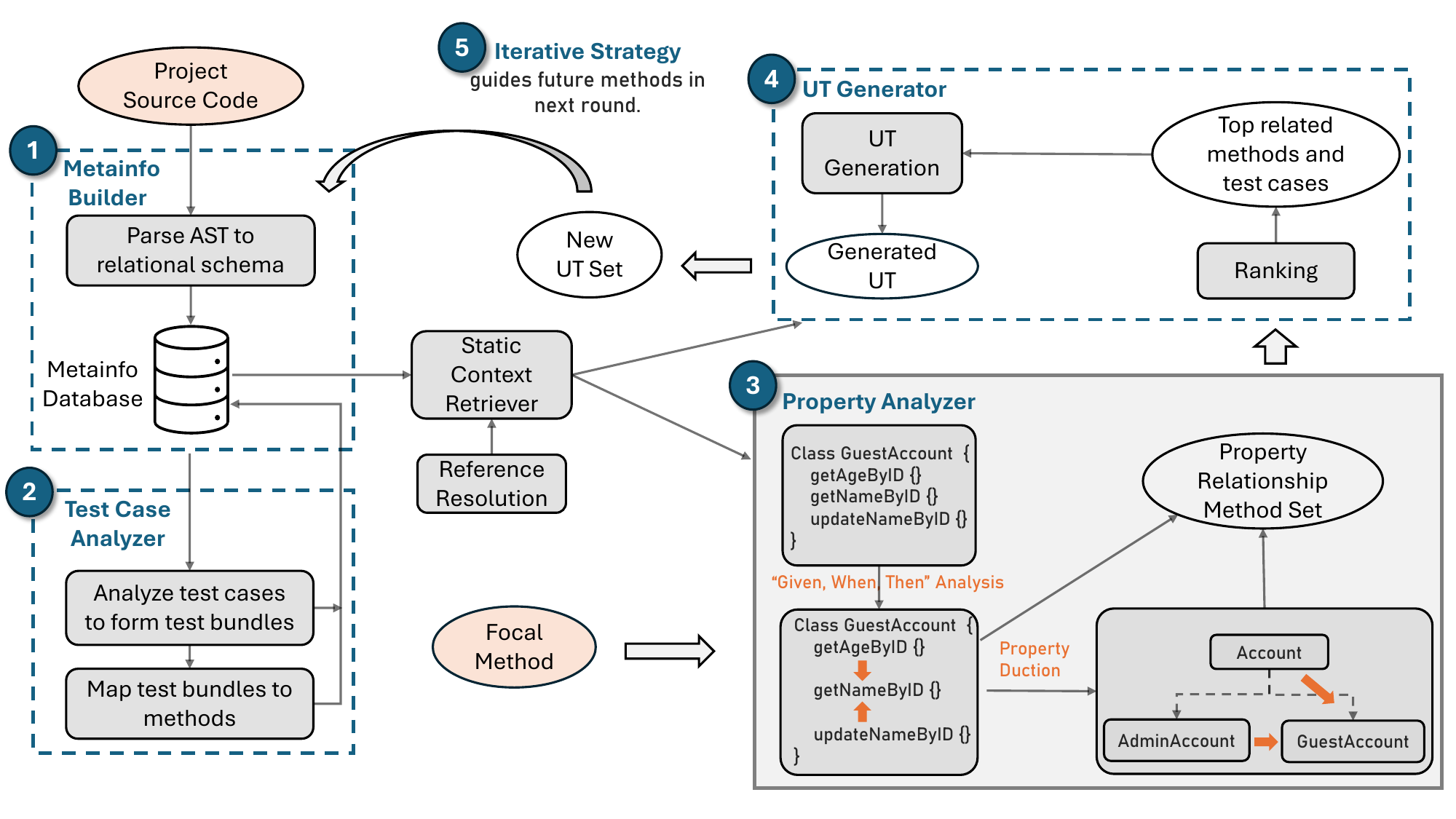}
    \vspace{-6pt}
    \caption{The overall workflow of \toolname.}
    \vspace{-6pt}
    \label{fig:workflow}
\end{figure}

First, the input project is processed by the \texttt{Metainfo Builder}, which parses the source code into an Abstract Syntax Tree (AST) and transforms it into a relational structure based on predefined schemas (e.g., Class, Method, Package). This structure is stored in the \textit{Metainfo Database}, ensuring efficient data retrieval for the following stages.
Next, existing test cases in the project are analyzed by the \texttt{Test Case Analyzer}. This process abstracts the test cases into \textit{Test Bundles}, which are then mapped back to their corresponding methods.
Third, the \texttt{Property Analyzer} takes the focal method and its class as input and performs reference resolution through static context retrieval. It conducts a ``Given, When, Then'' analysis to identify intra-class properties and uses property deduction to retrieve related methods across the entire repository. These results form the \textit{Property Relationship Method Set}.
Last, the \texttt{UT Generator} ranks the methods in the \textit{Property Relationship Method Set}, retrieves their corresponding test cases, and generates unit tests. The generated tests are added to the \texttt{UT Set}. 
After generating tests for all focal methods, the \texttt{Iterative Strategy} is applied. 
The newly generated tests are treated as historical tests for the next iteration to generate new unit tests. 



\subsection{Preliminaries}
In the context of property relationship analysis and unit test generation, there are three critical challenges to address:

\begin{itemize}[leftmargin=*]
    \item \textbf{Where to find?} In property relationship analysis and unit test generation, relevant code entities such as methods, classes, and interfaces must be efficiently located across large codebases, much like querying a relational database.
    \item \textbf{What to look for?} To understand a code snippet fully, especially for unit test generation, it’s necessary to resolve references to external entities (e.g., methods, variables, or classes) that may not be defined locally in the code under analysis.
    \item \textbf{What information to provide?} In scenarios where the external information is too large to fit within the context window of the language model (LLM), we need to prioritize and compress the relevant code data while preserving essential context.
\end{itemize}

To address these challenges, our methodology incorporates two key processes: \textit{Metainfo Extraction} and \textit{Static Context Retrieval}, both of which are essential for ensuring that the LLM has the necessary code structure and context for accurate analysis and test case generation.

\subsubsection{Metainfo Extraction}

\begin{wrapfigure}{r}{0.55\textwidth}
\vspace{-14pt}
    \centering
    \begin{lstlisting}[language=json, basicstyle=\scriptsize, numbers=left, numbersep=2pt, breaklines=true, postbreak=\mbox{\textcolor{red}{$\hookrightarrow$}\space}]
{
  "uri": "src/main/java/org/redisson/transaction/
  operation/set/MoveOperation.java.MoveOperation",
  "name": "MoveOperation",
  "file_path": "src/main/java/org/redisson/transaction/
  operation/set/MoveOperation.java",
  "superclasses": "SetOperation",
  "methods": [
    "[void]commit(CommandAsyncExecutor)",
    "[void]rollback(CommandAsyncExecutor)"
  ],
  "class_docstring": "@author Nikita Koksharov",
  "original_string": "", // Omitted
  "super_interfaces": [],
  "fields": [
    {
      "docstring": "",
      "modifiers": "private",
      "marker_annotations": [],
      "type": "String",
      "name": "destinationName"
    }
  ]
}
    \end{lstlisting}
    \caption{Simplified structured metadata for the \texttt{MoveOperation} class}
    \vspace{-10pt}
    \label{fig:move_operation}
\end{wrapfigure}


The \texttt{Metainfo Builder} processes the input project by parsing the source code into an Abstract Syntax Tree (AST) and transforming it into a relational structure based on predefined schemas (e.g., Class, Method, Package). The extracted data is stored in the \textit{Metainfo Database}, allowing for efficient retrieval of relevant information.

Figure \ref{fig:move_operation} illustrates structured metadata for the \texttt{MoveOperation} class, including details like its URI, methods, fields, superclasses, and additional documentation. Such structured organization of code data enables effective retrieval and supports tasks like property relationship analysis and test case generation. We also design schemas for other code entities, such as test classes, interfaces, abstract classes, and records.


\subsubsection{Reference Resolution Based on Scope Graph}
\label{sec:scope_graph}

In addition to storing the code structure, resolving references to external code entities is essential for understanding and generating accurate test cases. Our approach, \textit{Reference Resolution Based on Scope Graph}, addresses the challenge of determining ``What to look for?'' by identifying external variables, functions, or classes required to fully comprehend a code snippet. This process enhances the LLM's ability to analyze code by completing the context necessary for code understanding.

Traditional reference resolution often depends on language-specific libraries, such as RepoFuse~\cite{liang2024repofuserepositorylevelcodecompletion}, which relies on tools like jedi~\cite{jedi} for Python auto-completion. These approaches, while effective, are limited in scope, supporting only specific languages. The Language Server Protocol (LSP) \cite{LSP} improves on this by providing cross-language support for features like go-to-definition and find-references, but it introduces significant overhead due to its reliance on language parsers, making it slow and resource-intensive. Given these constraints, a lightweight and language-agnostic solution is essential for supporting large-scale codebases. 

To solve this problem, \toolname extends the capabilities of Scope Graphs \cite{neron2015theory}, which originally focus on resolving variable definitions within a single scope. 
While traditional scope graphs struggle with cross-class references, \toolname enhances this by identifying unresolved references across classes and files, completing lookups using the \textit{Metainfo Database}. 


\paragraph{\textit{Definition.}}
A \textbf{Scope Graph} \( G = (N, E) \) provides a structured representation of lexical scopes, declarations, and references in the program. 

\begin{itemize}
    \item Nodes \( N \) represent entities such as:
    \begin{itemize}
        \item \texttt{LocalScope}: A lexical scope within the code.
        \item \texttt{LocalDef}: A local definition (e.g., variable, function, or class).
        \item \texttt{LocalImport}: An import statement introducing external symbols.
        \item \texttt{Reference}: A usage of a symbol that must be resolved to its definition or import.
    \end{itemize}
    
    \item Edges \( E \) describe relationships between these entities:
    \begin{itemize}
        \item \texttt{ScopeToScope}: A parent-child relationship between scopes.
        \item \texttt{DefToScope}: Links a definition to the scope it belongs to.
        \item \texttt{ImportToScope}: Connects imports to the target scope.
        \item \texttt{RefToDef}, \texttt{RefToImport}: Capture references to definitions or imports.
    \end{itemize}
\end{itemize}

\paragraph{Graph Construction and Reference Resolution}
\toolname construct graphs by the following steps:

\begin{enumerate}[leftmargin=*]
    \item \textbf{Scope Insertion:} Each lexical scope is inserted into the graph, with its parent-child relationships captured using \texttt{ScopeToScope} edges.
    \item \textbf{Declaration Insertion:} For every symbol declared (e.g., variable or function), a \texttt{LocalDef} node is added and linked to the appropriate scope.
    \item \textbf{Reference Resolution:} When a symbol is referenced, the graph is traversed to find its corresponding definition or import via \texttt{RefToDef} or \texttt{RefToImport} edges. Unresolved references are flagged if no match is found in the current or parent scopes.
\end{enumerate}

\paragraph{Algorithm for Reference Resolution}
\toolname resolve references \( \mathit{Ref} \) within a code block \( cb \) via: 

\begin{enumerate}[leftmargin=*]
    \item \textbf{Reference Insertion:} For each reference \( r \) in the \( cb \), associate it with its lexical scope \( s \) and insert it into the Scope Graph:
    \[
    \text{insertRef}(r, s)
    \]
    

    \item \textbf{Declaration Lookup:} For each reference \( r \), search for a matching declaration in the current scope \( s \). If none is found, traverse the graph through \texttt{ScopeToScope} edges to the parent scopes, recursively searching for the definition:
    \[
    \text{findDecl}(r, s) \quad \text{(traverse through parent scopes)}
    \]
    If no declaration is found in the entire scope chain, extend the search via \texttt{RefToImport} edges to check for imports:
    \[
    \text{findDecl}(r, s) \quad \text{(via imports)}
\]

    \item \textbf{Unresolved Reference Detection:} If no matching declaration is found, mark the reference as unresolved and collect all such unresolved references: 
    \[
    UR = \{ r \in \mathit{Ref} \mid \text{resolve}(r) = \emptyset \}
    \]

    \item \textbf{Unresolved Reference Retrieval from Metainfo Database:} For each unresolved reference \( r \in UR \), \toolname queries the \textit{Metainfo Database} to retrieve its definition from other classes or files and aggregates the results in \( \text{resolveRef}(cb) \):
    \[
    \text{resolveRef}(cb) = \{ \text{ QuerymMetainfo}(r) \mid r \in UR \}
    \]
\end{enumerate}

\subsubsection{AST-Based Code Compression}
When retrieving relevant references for a code snippet, it is crucial to provide only the necessary information to fit within the LLM's context window. Hence, \toolname uses two mechanisms tailored for different scenarios: \textit{Class Montage} and \textit{Class Shrink}.

\paragraph{\textit{Class Montage}}
When analyzing test cases, it is necessary to map test methods to their corresponding target methods.
Specifically, given a test method, it needs to 
trace through the call chain to find the final target function. In such cases, knowing just the method signatures of the class is sufficient.
Additionally, when understanding code, it is often enough to know what a referenced class does without needing all the implementation details.
Here, a simplified ``montage'' of the class is sufficient for certain scenarios. 
\textit{Class Montage} abstracts implementation details by including only method signatures, fields, and inner classes while omitting method bodies. For example, in Figure \ref{fig:class_montage_of_hashkey}, the \texttt{RedisCommonBatchExecutor} class is reduced to its signatures and fields, offering a compact representation for LLM processing.

\paragraph{\textit{Class Shrink}} 
After retrieving the \textit{Property Relationship Method Set} \( \mathit{R}(m_t) \) through \textit{property-based retrieval}, providing the entire class to the LLM can be inefficient due to input length limitations and the risk of including irrelevant details. To address this, \toolname employs \textit{Class Shrink}, which focuses on keeping only the methods in \( \mathit{R}(m_t) \) and fields of the class, discarding everything unrelated. For example, as shown in Figure \ref{fig:class_shrink_of_hashkey}, in the \texttt{RedisCommonBatchExecutor} class \footnote{\url{https://github.com/redisson/redisson/blob/master/redisson/src/main/java/org/redisson/command/CommandBatchService.java}}, only focal method \texttt{retryAttempts} and relevant method \texttt{retryInterval} are kept, while unrelated methods like \texttt{sendCommand} are discarded. This ensures that the LLM focuses on the relevant context without unnecessary information.


\begin{figure}[h]
\vspace{-6pt}
    \centering
    \begin{subfigure}[b]{0.49\textwidth}
        \centering
        \begin{lstlisting}[language=java, basicstyle=\scriptsize, breaklines=true, numbers=left, numbersep=2pt, postbreak=\mbox{\textcolor{red}{$\hookrightarrow$}\space}]
public class RedisCommonBatchExecutor {
    private final Entry entry;
    private final BatchOptions options;
    private static int timeout(ConnectionManager connectionManager, BatchOptions options);
    private static int retryInterval(ConnectionManager connectionManager, BatchOptions options);
    private static int retryAttempts(ConnectionManager connectionManager, BatchOptions options);
    @Override
    protected void sendCommand(CompletableFuture<Void> attemptPromise, RedisConnection connection);
}
        \end{lstlisting}
        \caption{\texttt{RedisCommonBatchExecutor}'s Class Montage}
        \label{fig:class_montage_of_hashkey}
    \end{subfigure}
    \hfill
    \begin{subfigure}[b]{0.48\textwidth}
        \centering
        \begin{lstlisting}[language=java, basicstyle=\scriptsize, breaklines=true, numbers=left, numbersep=2pt, postbreak=\mbox{\textcolor{red}{$\hookrightarrow$}\space}]
public class RedisCommonBatchExecutor {
    private final Entry entry;
    public RedisCommonBatchExecutor() {
        // Constructor
    } 
    
    private static int retryInterval(ConnectionManager connectionManager, BatchOptions options) {
        // Implementation omitted
    }
    
    private static int retryAttempts(ConnectionManager connectionManager, BatchOptions options) {
        // Implementation omitted
    }
}
        \end{lstlisting}
        \caption{\texttt{RedisCommonBatchExecutor}'s Class Shrink}
        \label{fig:class_shrink_of_hashkey}
    \end{subfigure}
    \vspace{-6pt}
\caption{Class Montage and Class Shrink of \texttt{RedisCommonBatchExecutor}: a comparative example illustrating two mechanisms for reducing class information. The left side shows Class Montage, which abstracts the class by including its signature and fields. The right side shows Class Shrink, which reduces the class by retaining only the methods and fields relevant to the focal method.}
\vspace{-12pt}
 \label{fig:combined_class_montage_shrink}
\end{figure}

\subsection{Test Case Analysis} 
When a relevant test case is identified and deemed suitable for reference, it is essential to provide not just the test case itself but also the full context in which it operates. This context includes the test case itself, its associated fixtures, any referenced variables (whether within the class or from external sources), mock objects, or other forms of dependencies.

\paragraph{Definition}
Let \( c \) be a test class containing \( n \) test cases, \( m \) fixtures, as well as several imported modules, class variables, and functions. The test class \( c \) is formally defined as the combination of its test cases, fixtures, imports, and class members:
\[
c = \{ \text{tc}_1, \text{tc}_2, \dots, \text{tc}_n \} + \{ \text{Fixture}_1, \text{Fixture}_2, \dots, \text{Fixture}_m \} + \text{Imports}(c) + \text{ClassMembers}(c)
\]


\paragraph{Definition}
Similarly, the Test Bundle Set \( \text{TB}_r \) for a specific method \( m_r \) is defined as the collection of all test bundles related to \( m_r \). Each test bundle \( \text{tb}(c, i) \) for a test case \( \text{tc}_i \) in test class \( c \) is defined as the combination of the test case, its related fixtures, imports, and class members:
\begin{align*}
\text{tb}(c, i) = & \, \text{tc}_i + \{ \text{Fixture}_j \mid \text{Fixture}_j \text{ is used by } \text{tc}_i \} \\
& + \{ \text{Imports}(c) \mid \text{Imports}(c) \text{ is used by } \text{tc}_i \} \\
& + \{ \text{ClassMembers}(c) \mid \text{ClassMembers}(c) \text{ is used by } \text{tc}_i \}
\end{align*}
The set of test bundles is denoted as \( \text{TB} \), including all \( \text{tb}(c, i) \) for every test class \( c \) and test case \( \text{tc}_i \).




\begin{wrapfigure}{r}{0.52\textwidth}
\vspace{-20pt}
    \centering
    \begin{lstlisting}[language=json, basicstyle=\scriptsize, numbers=left, numbersep=2pt, breaklines=true, postbreak=\mbox{\textcolor{red}{$\hookrightarrow$}\space}]
{
    "file_path": "src/test/.../JCacheTest.java",
    "name": "JCacheTest",
    "dependencies": [...],
    "class_members": {
      "variables": [...],
      "methods": [],
      "nested_classes": ["ExpiredListener"]
    },
    "fixtures": ["beforeEach"],
    "test_cases": [{
        "name": "testClose",
        "primary_tested": ["Cache.close()"],
        "external_dependencies": {...},
        "fixtures_used": ["beforeEach"],
        "project_specific_resources":
        ["TestUtil.logTestResult(String, int)"]
      }],
    ...
}
    \end{lstlisting}
    \vspace{-12pt}
    \caption{Simplified test case analysis result for \texttt{JCacheTest}}
    \vspace{-12pt}
    \label{fig:jcache_test_metadata}
\end{wrapfigure}

The test case analysis process consists of two main steps. First, it performs test bundle abstraction, using static context retrieval and LLM-based analysis to generate the test bundles. Each test bundle encapsulates the relevant context for a given test case. Then, the test bundles are mapped to the corresponding methods based on the analysis results. 

Specifically, the input to the test bundle abstraction includes the test class and the \textit{Class Montage} of the class being tested. The \textit{Class Montage} helps the LLM identify the precise signature of the methods under test. This input is fed into the LLM, and the output is a structured JSON format, as shown in Figure \ref{fig:jcache_test_metadata}. The ``test\_cases'' field in the JSON stores the individual test bundles, each containing details of which method the test case targets, with its full context, including external dependencies, fixtures used, and project-specific resources.

This approach ensures that all relevant contexts, dependencies, and configurations influencing the test's behavior are fully accounted. By leveraging the \textit{Test Bundle}, the use of test cases becomes more precise and context-aware.





\subsection{Property Retrieval}

\label{sec:property_analysis_and_deduction}

To address the \textit{property retrieval} challenge in section \ref{sec:property_retrieval_challenge}, we first define the property and then present the analysis of property relationships within a class and across classes.

\subsubsection{Property Definition}
\label{sec:property_definition}


A property relationship between two methods \( m_A \) and \( m_B \) exists if \( m_A \) can provide reference information that aids in the test case generation for \( m_B \) in any one or more of the three phases: \emph{Given}, \emph{When}, and \emph{Then}.
The property relationship \( P(m_A, m_B) \) is defined as:
\[
P(m_A, m_B) \iff P_{\textit{Given}}(m_A, m_B) \lor P_{\textit{When}}(m_A, m_B) \lor P_{\textit{Then}}(m_A, m_B)
\]


The relations in \emph{\textbf{Given}} phase involve similar preconditions, object initialization, setup, etc: 
\[
P_{\textit{Given}}(m_A, m_B) \iff \text{The preconditions or setup of } m_A \text{ can be reused or referenced for } m_B
\]

The relation in \emph{\textbf{When}} phase refers to the similarities in terms of 
call sequences, parameters, dependencies, etc, which is defined as: 
\[
P_{\textit{When}}(m_A, m_B) \iff \text{The method invocation patterns or dependencies of } m_A \text{ apply to } m_B
\]

The relations in \emph{\textbf{Then}} phase involves similar asserting the expected results: 
\[
P_{\textit{Then}}(m_A, m_B) \iff \text{The assertions or exception handlings from } m_A \text{ are relevant for } m_B
\]



In cases where all three conditions are met:
\[
P_{\textit{Complete}}(m_A, m_B) \iff P_{\textit{Given}}(m_A, m_B) \land P_{\textit{When}}(m_A, m_B) \land P_{\textit{Then}}(m_A, m_B)
\]


\subsubsection{Class-Level Property Analysis (Intra-Class)}
\label{sec:intra_property_analysis}









\emph{Class-Level Property Analysis} focuses on identifying property relationships between methods within the same class, forming the foundation for extending the analysis to cross-class relationships.
Given a focal method \( m_t \) in class \( c_t \), the analysis examines other methods within the class \( M(c_t) = \{ m_1, m_2, \dots, m_n \} \), including any inherited methods from parent classes, which are recursively incorporated via the \textit{MetaInfo Database}. The goal is to find methods \( m_r \) with property relationships to \( m_t \) across the \emph{Given}, \emph{When}, and \emph{Then}.

\begin{wrapfigure}{r}{0.50\textwidth}
    \vspace{-15pt}
    \centering
    \includegraphics[width=0.50\textwidth]{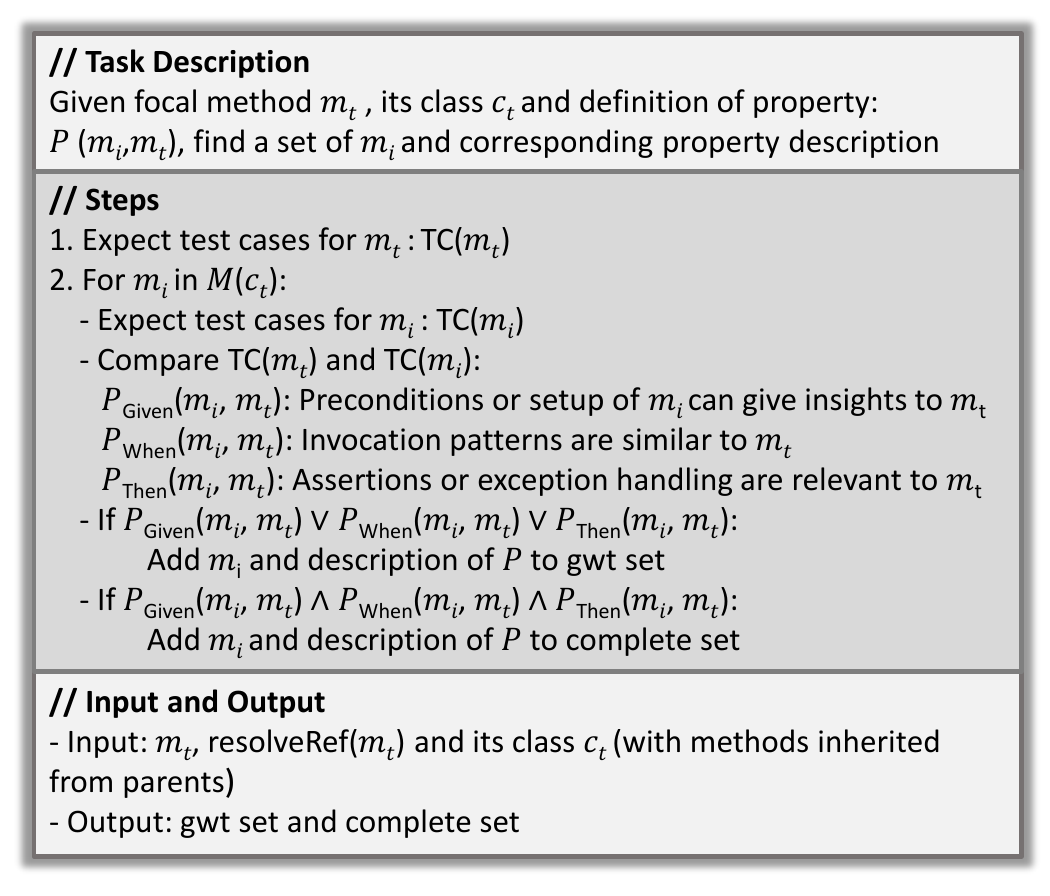}
    \vspace{-25pt}
    \caption{Formal version of the prompt for analyzing properties.}
    \label{fig:property_analyzer_prompt}
    \vspace{-15pt}
\end{wrapfigure}

As illustrated in Figure~\ref{fig:property_analyzer_prompt}, the formal version of the prompt for property analysis is provided. First, in the ``Task Description'' section, we define the task for the LLM, which includes the property definition and instructing the LLM to find the set of methods that have property relationships with the focal method \( m_t \).
The core of the prompt lies in leveraging \emph{Rule-based Reasoning} \cite{servantez2024chain} and \emph{Chain of Thought (CoT) reasoning}  \cite{chen2023you}, as shown in the ``Steps'' section of Figure~\ref{fig:property_analyzer_prompt}. Instead of directly identifying all property relationships for a method \( m_t \), which can overwhelm LLM due to complex inter-method dependencies and relationships, causing it to lose focus, we decompose the problem into smaller manageable parts and let LLM solve it step by step.

Specifically, the LLM is first instructed to imagine a set of expected test cases \( \mathit{TC}(m_t) \) for the focal method \( m_t \) by reasoning. Then, it traverses all methods \( m_i \) in the class \( c_t \). For each method \( m_i \), the LLM imagines the test cases \( \mathit{TC}(m_i) \), compares \( \mathit{TC}(m_t) \) with \( \mathit{TC}(m_i) \), and identifies potential property relationships in the \emph{Given}, \emph{When}, and \emph{Then} phases. If a relationship is found in any phase, the method \( m_i \) and its corresponding relationship description are added to the \emph{gwt set}. If relationships are found in all three phases, 
they are added to the \emph{complete set}.
Finally, the input to the LLM consists of the focal method \( m_t \), the external reference information retrieved through \texttt{resolveRef(\( m_t \))}, and the source code of class \( c_t \), including any inherited methods from parents. The LLM is tasked with outputting the \emph{gwt set} and \emph{complete set} as structured JSON.



The generated JSON output provides a clear mapping of property relationships between methods. As shown in Table \ref{tab:removeFirst_enhancements}, it highlights the methods that share property relationships with \texttt{removeFirst()}, along with corresponding descriptions and confidence scores. Specifically, the methods \texttt{poll()}, \texttt{remove()}, and \texttt{getFirst()} offer valuable references to \texttt{removeFirst()} across all three phases \emph{Given}, \emph{When}, and \emph{Then}—thus they are categorized under \textbf{Complete}. On the other hand, methods such as \texttt{initializeQueue()}, \texttt{validateQueueState()}, and \texttt{verifyRemoval()} provide specific references in the \emph{Given}, \emph{When}, and \emph{Then} phases respectively.

Furthermore, Table \ref{tab:removeFirst_enhancements} also shows their respective confidence scores with the focal method. These scores represent the confidence in each relationship, with higher scores (such as \texttt{poll()} at 0.90) indicating a closer functional similarity to \texttt{removeFirst()}. Lower scores, like the 0.70 for \texttt{validateQueueState()}, suggest a more indirect but still relevant relationship, such as ensuring the queue is not empty before calling \texttt{removeFirst()}.

\begin{table}[thbp]
\centering
\footnotesize
\vspace{-10pt}
\caption{Property Relationships and Enhancements for \texttt{removeFirst()}}
\vspace{-10pt}
\begin{tabularx}{\textwidth}{|l|l|X|c|c|}
\hline
\textbf{Phase} & \textbf{Method Name}          & \textbf{Reason}                                                                & \textbf{Confidence} & \textbf{External} \\ \hline
\multirow{3}{*}{\textbf{Complete}} 
               & \texttt{poll()}               & Similar control flow for removing the first element asynchronously.            & 0.90                & No                \\ \cline{2-5} 
               & \texttt{remove()}             & Internally calls \texttt{removeFirst()}, making both interchangeable.          & 0.85                & No                \\ \cline{2-5} 
               & \texttt{getFirst()}           & Accesses the first element but does not remove it.                             & 0.80                & No                \\ \hline
\textbf{Given} & \texttt{initializeQueue()}    & Queue initialization is needed to ensure proper configuration.                 & 0.80                & No                \\ \hline
\textbf{When}  & \texttt{validateQueueState()} & Ensures the queue is not empty before calling \texttt{removeFirst()}.          & 0.70                & No                \\ \hline
\textbf{Then}  & \texttt{verifyRemoval()}      & Confirm the first element was successfully removed from the queue.            & 0.85                & No                \\ \hline
\end{tabularx}
\vspace{-10pt}
\label{tab:removeFirst_enhancements}
\end{table}

It is important to note that the relevance of property relationships is not confined to methods with semantic similarity. For instance, even methods that perform semantically opposite tasks, such as \emph{encode} and \emph{decode}, can share a common test case structure across all three phases (\emph{Given}, \emph{When}, and \emph{Then}) and thus have property relationships that can enhance test case generation.

In a word, this approach breaks down the task into three more manageable components, each focusing on one of the three phases of test case generation. By guiding the LLM to explore each phase separately, \toolname enables it to incrementally uncover property relationships.






\subsubsection{Repository-Level Property Analysis (Inter-Class)}
\label{sec:repo_level_property_analysis}

\emph{Repository-Level Property Analysis} extends property analysis across class boundaries by considering relationships such as inheritance, sibling classes, and common interface implementations. These relationships introduce opportunities for cross-class property relationships, where test cases for methods in one class can provide reference information for generating test cases for methods in another class. 


\paragraph{General Property Relationship Formula}

For any two classes \( c_1 \) and \( c_2 \) related by inheritance, sibling class, or common interface implementation, and for two methods \( m_{A} \in M(c_1) \) and \( m_B \in M(c_1) \), the general condition for extending property relationships across classes is as follows:
\[
P(m_A \in c_1, m_B \in c_1) \land m_A \in M_{\text{shared}} \implies 
    P(m_A \in c_2, m_B \in c_1)
\]
This formula means that if a method \( m_A \in c_1 \) enhances method \( m_B \in c_1 \), and \( m_A \) belongs to the shared methods \( M_{\text{shared}} \) between $c_1$ and $c_2$, then \( m_A \) from \( c_2 \) can also enhance \( m_B \) in \( c_1 \). This provides a unified way to express how shared methods across related classes contribute to test generation.

\paragraph{Inheritance Relationships}

For a parent class \( c_p \) and its child class \( c_c \),

\begin{itemize}[leftmargin=*]
    \item \( M_{\text{shared}} = M(c_p) \cap M(c_c) \) represent shared method set between the parent and child classes.
\end{itemize}

Shared methods between parent and child classes can enhance methods in either class. If a shared method \( m_A \in M_{\text{shared}} \) enhances \( m_B \in c_c \), then the corresponding method in \( c_p \) can similarly enhance \( m_B \), and vice versa. This bidirectional relationship follows from the general property relationship formula and enables mutual test case enhancement across the inheritance hierarchy.

\paragraph{Sibling Class Relationships}

For sibling classes \( c_1 \) and \( c_2 \), which share a common parent class,

\begin{itemize}[leftmargin=*]
    \item \( M_{\text{shared}} = M(c_1) \cap M(c_2) \) represent shared method set between these sibling classes.
\end{itemize}

If \( P(m_A \in c_1, m_B \in c_1) \) holds for a shared method \( m_A \in M_{\text{shared}} \), then \( m_A \in c_2 \) can enhance \( m_B \in c_1 \), following the general formula.

\paragraph{Common Interface Implementations}

For classes \( c_1 \) and \( c_2 \) that implement the same interface \( I \), let:

\begin{itemize}[leftmargin=*]
    \item \( M_{\text{shared}} = M(c_1) \cap M(c_2) \cap M(I) \) represent method set shared by both classes and the interface.
\end{itemize}

If \( P(m_A \in c_1, m_B \in c_1) \) holds for a shared method \( m_A \), then the corresponding method \( m_A \in c_2 \) can also enhance \( m_B \in c_1 \), as per the general formula.

\paragraph{Consolidation}

In all cases, the enhancement of property relationships depends on shared methods \( M_{\text{shared}} \). If \( m_A \in M_{\text{shared}} \) enhances \( m_B \) in one class, the same method \( m_A \) from a related class can enhance \( m_B \) as well, ensuring property relationships extend across class boundaries and support more robust test generation.

Additionally, in section \ref{sec:intra_property_analysis}, when the LLM identifies methods from external classes that can be directly reused (e.g., when the focal method calls an external function in a substitutive manner), the ``External'' field in JSON is included to indicate whether the related method comes from an external class, and this also facilitates inter-class property analysis.

\subsection{Iterative Test Case Generation}
In this subsection, we introduce the approach to generating effective test cases for the focal method, solving \textit{Generating Effective Unit Tests} challenge in section \ref{sec:ut_generation_challenge}.
\subsubsection{Property-Based Retrieval and Ranking}

Here, we explain the process of generating test cases for the focal method \( m_t \) by leveraging property relationships and a ranking strategy.

\begin{algorithm}[!th]
\caption{Test Case Generation for a Focal Method}
\footnotesize
\label{alg:test_generation_focal_method}
\SetAlgoNoEnd 
\LinesNumbered
\KwIn{Focal method: $m_t$; Property relation method set: $R(m_t)$; Existing test bundle set: $TB$; 
Property-based retrieval prompt: \textit{property\_prompt}; Fallback prompt: \textit{fallback\_prompt};  
}
\KwOut{Generated unit test $ut_t$ for $m_t$}

\SetKwFunction{GetTestCases}{GetTestCases}
\SetKwFunction{RankMethods}{RankMethods}
\SetKwFunction{MinimizeTestBundle}{MinimizeTestBundle}
\SetKwFunction{GetStaticContext}{GetStaticContext}
\SetKwFunction{QueryMetainfo}{QueryMetainfo}
\SetKwFunction{ClassShrink}{ClassShrink}
\SetKwFunction{CallLLM}{CallLLM}
\SetKwFunction{ResolveRef}{ResolveRef}
\SetKwFunction{GetUnresolvedRef}{GetUnresolvedRef}

/* \textit{Initialization} */ \\
$S_{\textit{origin}} \leftarrow \emptyset  $\hspace{1em}\tcp{Store related methods with test bundles and their property relationship to $m_t$}
$S_{\textit{ranked}} \leftarrow \emptyset$\hspace{1em}\tcp{Store ranked methods with test bundles and their property relationship to $m_t$}
$D \leftarrow \{\}$\hspace{1em}\tcp{Dictionary (ranked method $\rightarrow$ test bundles)}
$\textit{SC} \leftarrow \emptyset$\hspace{1em}\tcp{Static context}


\vspace{1em}

/* \textit{Rank} */ \\
\ForEach{$m_r \in R(m_t)$}{\label{line:filter_start}
    \If{$TC_r \neq \emptyset$}{
        $S_{\textit{origin}} \leftarrow S_{\textit{origin}} \cup \{ m_r \}$\label{line:filter_end}\;
    }
}

$S_{\textit{ranked}} \leftarrow \texttt{rank}(S_{\textit{origin}})$ \label{line:rank}
// \textit{intra-class complete} $>$ \textit{intra-class GWT} $>$ \textit{inter-class complete} $>$ \textit{inter-class GWT}

\vspace{1em}

/* \textit{Pack context} */ \\
$SC \leftarrow SC \cup \ResolveRef(m_t)$;\hspace{1em}\tcp{Resolve reference for $m_t$}  \label{line:get_ur}

\If{$S_{\text{ranked}} \neq \emptyset$}{
$S_{\text{intra}}, S_{\text{inter}} \leftarrow$ Split $S_{\textit{ranked}}$ into intra-class and inter-class methods\; \label{line:split}
\If{$S_{\text{intra}} \neq \emptyset$}{ \label{line:intra_start}
    $SC \leftarrow SC \cup \ClassShrink(m_t, S_{\text{intra}})$\;
} \label{line:intra_end}
\ForEach{$m_r \in S_{\text{inter}}$}{\label{line:inter_start}
    $SC \leftarrow SC \cup \QueryMetainfo(m_r)$\;
}\label{line:inter_end}

\ForEach{$m_r \in S_{\textit{ranked}}$}
{\label{line:minimize_start}
    $D \leftarrow D \cup \{ m_r \rightarrow TB_r \}$\;
}

$D_m \leftarrow \MinimizeTestBundle(D);$\hspace{1em}\tcp{Remove redundancy}
\label{line:minimize_end}

$I \leftarrow \textit{property\_prompt} + {m_t} + SC + S_{\text{ranked}} + D_m$\;
}
\Else{\label{line:fallback_start}
    $I \leftarrow {m_t} + SC + \textit{fallback\_prompt}$; \label{line:fallback_end}
} 

\vspace{1em}
$ut_t \leftarrow \CallLLM(I)$ \label{line:call_llm}
\end{algorithm}

Algorithm~\ref{alg:test_generation_focal_method} outlines the process for generating a unit test. The inputs are the focal method $m_t$, the property relation method set $R(m_t)$ (from the \textit{property analysis}), and the test bundle set $TB$.

\toolname first obtains methods with test cases in $R(m_t)$ to create $S_{\textit{origin}}$ (line \ref{line:filter_start} - \ref{line:filter_end}), then ranks them to form $S_{\textit{ranked}}$ (line\ref{line:rank}). It prioritizes methods at the class level over those at the repository level. Within each level, \textit{complete} relationships are selected first, with up to $N$ methods chosen. Afterward, for the \emph{Given}, \emph{When}, and \emph{Then} phases, \toolname selects up to $N$ methods per phase, ensuring at least one per phase if available. 
The final ranking order is: \textit{class-level complete} > \textit{class-level GWT} > \textit{repo-level complete} > \textit{repo-level GWT}, with a maximum of $4N$ methods selected for test generation. The default $N$ is three, but it can be adjusted by the user.

After ranking, \toolname begins packing the context.
First, based on \textit{Reference Resolution Using Scope Graph}, \toolname call \( \text{resolveRef}(m_t) \) to resolve references, ensuring that all external references used by $m_t$ are resolved (line \ref{line:get_ur}). 
Next, $S_{\textit{ranked}}$ is split into intra-class methods $S_{\text{intra}}$ and inter-class methods $S_{\text{inter}}$ (line \ref{line:split}). 
For $S_{\text{intra}}$, \emph{Class Shrink} is applied to remove redundancy, and the resulting context is added to $SC$ (lines \ref{line:intra_start}-\ref{line:intra_end}). 
For each method $m_r$ in $S_{\text{inter}}$, \toolname queries the metainfo database (\texttt{QueryMetainfo}) to get the detailed information of $m_r$ and add them to $SC$ (lines \ref{line:inter_start}-\ref{line:inter_end}).
Simultaneously, \toolname gets all the test bundle of methods in $S_{ranked}$, storing them in $D$ and minimizes $D$ to $D_m$ to reduce redundancy (lines \ref{line:minimize_start}-\ref{line:minimize_end}). The final input to the LLM, $I$, consists of the property prompt, $m_t$, $SC$, $S_{\textit{ranked}}$, and $D_m$.
If $S_{\textit{ranked}}$ is empty, a fallback strategy is applied to generate test cases directly for the focal method (lines \ref{line:fallback_start}-\ref{line:fallback_end}). In this case, the final input to the LLM is $m_t$, $SC$, along with the fallback prompt. The \texttt{CallLLM} function is invoked with $I$, and the generated unit test $ut_t$ is returned (line \ref{line:call_llm}).

Following previous work \cite{liu2024llm, gu2024testart}, \toolname incorporates a repair mechanism where test cases generated in the first round are corrected in subsequent iterations. The faulty test case, along with the focal method and its static context, is provided to the LLM for refinement in the next round.

\subsubsection{Iterative Strategy}

Having explained the approach to generating unit tests for a focal method, this section outlines the iterative strategy to optimize test generation across the entire project.

Given the set of all focal methods \( M \) in a project, the \textit{Iterative Test Case Generation Strategy} begins by generating test cases for methods 
\( m_A \in M \), if there exists a property relationship \( P(m_A, m_B) \) with a related method \( m_B \), where \( m_B \) already has an existing test case \( \text{tc}_B \), formalized as:
\[
\mathcal{C}(m_A, m_B) = P(m_A, m_B) \wedge TC_B \neq \emptyset
\]

During each iteration, 
the newly generated test cases \( \text{tc}_A \) are added to the overall test suite \( TC \). 
Once all applicable focal methods are processed in the current iteration, the newly generated test cases undergo \textit{Metainfo Extraction} and \textit{Test Case Analysis} to update the reference information. 
The process then continues to the next iteration, generating test cases for additional focal methods that satisfy \( \mathcal{C}(m_A, m_B) \). 
This iterative process continues, aiming to generate test cases for as many focal methods as possible using the Property-Based Retrieval Augmentation approach. Each round introduces new test cases that expand the number of focal methods covered until no further test cases can be generated for additional methods.

\section{EVALUATION}
We implement our tool \toolname using 8,000 lines of Python code, and we evaluate \toolname and answer the following
research questions:


\begin{itemize}[leftmargin=*]
    \item \textbf{(RQ1)}: How effective is our approach in generating unit tests for a given focal method in terms of correctness and completeness?
    \item \textbf{(RQ2)}: How well is the maintainability of the generated unit tests?
    \item \textbf{(RQ3)}: How does the \textit{iterative strategy} contribute to the performance of our approach?
    \item \textbf{(RQ4)}: What patterns exist in \emph{property relationships}, and what common characteristics can be identified among them?
\end{itemize}


\subsection{Overall Effectiveness}
\label{sec:rq1}
To demonstrate the overall effectiveness of our tool, we compare it against state-of-the-art (SOTA) solutions. Previous approaches tend to generate large batches of test cases for an entire project, but in real-world development, users are more likely to need unit tests generated for individual focal methods. Therefore, in our experiments, we adopt a strategy where tests are generated for each focal method independently, allowing us to better measure the effectiveness of the tool in more practical, user-centric scenarios.

\textbf{Dataset}
We utilize 4 moderately complex projects with >10,000 lines of code from the datasets provided by HITS \cite{wang2024hits} and ChatUnitest\cite{xie2023chatunitest}. 
We additionally crawl 8 repositories from GitHub, bringing the total number of projects to 12. The selection criteria followed those outlined in HITS \cite{wang2024hits}.
Projects were required to have at least 150 stars to ensure community interest and were updated within the last month to guarantee active maintenance. The dataset encompasses a variety of domains, including utilities, parsers (e.g., HTML or expression parsers), and network protocols.
Finally, we generate unit tests for 1,515 focal methods across 12 projects.



\textbf{Metrics.} Previous works typically generate all test cases in a single batch and then report the overall coverage for the entire project. 
Differently, our strategy is to input one focal method at a time and generate a test file specifically for the focal method. 
We run the generated tests separately and use \texttt{JaCoCo}~\footnote{\url{https://github.com/jacoco/jacoco}}  to calculate the coverage for the focal method.

The results of the execution are categorized as follows:
\begin{itemize}[leftmargin=*]
    \item \textbf{Compilation and Runtime Errors}: Compilation errors may arise from issues like undeclared variables, while runtime errors can include execution issues like uncaught exceptions.
    \item \textbf{Assertion Errors}: These refer to assertion errors where the expected value does not match the actual value. Note that, to measure the effectiveness of test generation tools, we assume the focal method is correctly implemented.
    \item \textbf{Successful Execution}: This means the generated test runs successfully. 
    More specifically, \textbf{Full Coverage}: that both line and branch coverage are 100\%.
\end{itemize}


\textbf{Baseline and Configuration}
Previous SOTA tools like \texttt{ChatUnitTest}~\cite{xie2023chatunitest} and \texttt{HITS} \cite{wang2024hits} have set benchmarks in unit test generation.
However, during our experiments, the current open-source implementation of \texttt{HITS} encounters significant runtime issues, making it unusable for generating unit tests at the method level. 
Hence, we opt to compare our tool primarily with \texttt{ChatUnitTest} and the baseline large language model (LLM) using traditional Retrieval-Augmented Generation (RAG). For the LLM, we use \texttt{DeepSeek-V2.5} \cite{zhu2024deepseek}, which offers an optimal balance between cost and performance. 
It is widely recognized and exceeds GPT-4 specifically in terms of coding capabilities \cite{ChatBotArena}. Additionally, we include EvoSuite as another baseline in our comparison. The specific experimental parameters are as follows:

\begin{itemize}[leftmargin=*]
\item For each LLM-based tool, before each run, we specify the focal method for which the tool generates unit tests,
with a maximum of two repair rounds allowed. This process is repeated multiple times.
Private methods and methods with only one line of effective code are filtered out. For methods with nested or anonymous inner classes, only the outer method is tested.
\item To ensure fairness, \toolname generates a test file for each focal method with \textit{N} test methods, similar to the baseline LLM. For \texttt{ChatUnitTest}, we use its default configuration to generate 5 test methods per focal method. The baseline LLM also receives the source code of the containing class as context (as the default RAG strategy), with the prompt adapted from ChatTester \cite{yuan2023no}, effective for Java unit test generation.
\item For EvoSuite~\cite{fraser2011evosuite}, we use its latest runtime version 1.0.6, which supports only Java 8 and JUnit 4. As a result, the number of methods testable by EvoSuite is smaller compared to other tools. Specifically, EvoSuite is used to generate tests for 389 methods.
Moreover, EvoSuite's search budget (time limit) is set to 120 seconds 
, with all other parameters left at their default settings.
\end{itemize}


\subsubsection{Test Generation Results}
\begin{table}[t]
    \centering
    \footnotesize
    \caption{Test Generation Results Across Tools}
    \vspace{-10pt}
    \label{tab:test_generation_results}
    \begin{tabular}{@{}lcccc@{}}
        \toprule
        \textbf{Metric} & \textbf{\toolname (Our Tool)} & \textbf{ChatUnitTest} & \textbf{LLM (with default RAG)} & \textbf{Evosuite} \\
        \midrule
        Compilation and Run Errors & \textbf{393 (25.9\%)} & 923 (60.9\%) & 1155 (76.2\%) & 232 (59.6\%) \\
        Assert Errors              & 210 (13.9\%) & 228 (15.0\%) & \textbf{127 (8.4\%)} & 53 (13.6\%) \\
        Successful Executions      & \textbf{912 (60.2\%)} & 364 (24.0\%) & 233 (15.4\%) & 104 (26.7\%) \\
        Full Coverage              & \textbf{821 (54.2\%)} & 323 (21.3\%) & 200 (13.2\%) & 99 (25.4\%) \\
        \hline
        \textbf{Total Methods Tested} & 1515 & 1515 & 1515 & 389 \\
        \bottomrule
    \end{tabular}
    \vspace{-10pt}
\end{table}
Table \ref{tab:test_generation_results} shows the results of evaluated tools. 
In terms of \textbf{correctness}, \toolname demonstrated significantly fewer compilation and run errors (25.9\%) compared to ChatUnitTest (60.9\%) and the LLM with default RAG (76.2\%). It also has the highest rate of successful executions at 60.2\%, outperforming ChatUnitTest (24.0\%), the LLM with default RAG (15.4\%), and EvoSuite (26.7\%). These results highlight the higher reliability of test cases generated by \toolname.
In terms of \textbf{completeness}, \toolname achieved full coverage in 54.2\% of cases, outperforming ChatUnitTest (21.3\%), the LLM with default RAG (13.2\%), and EvoSuite (25.4\%). This demonstrates \toolname’s superior ability to generate comprehensive tests that provide broader coverage of the code under test.

Overall, the results confirm that \toolname not only produces test cases with higher correctness by reducing errors and increasing successful executions but also ensures greater completeness by achieving full coverage in a significantly larger proportion of cases. These findings underscore the effectiveness of \toolname in generating reliable, high-coverage test cases. 



\begin{figure}[h]
    \centering
    \vspace{-10pt}
    \includegraphics[width=0.85\textwidth]{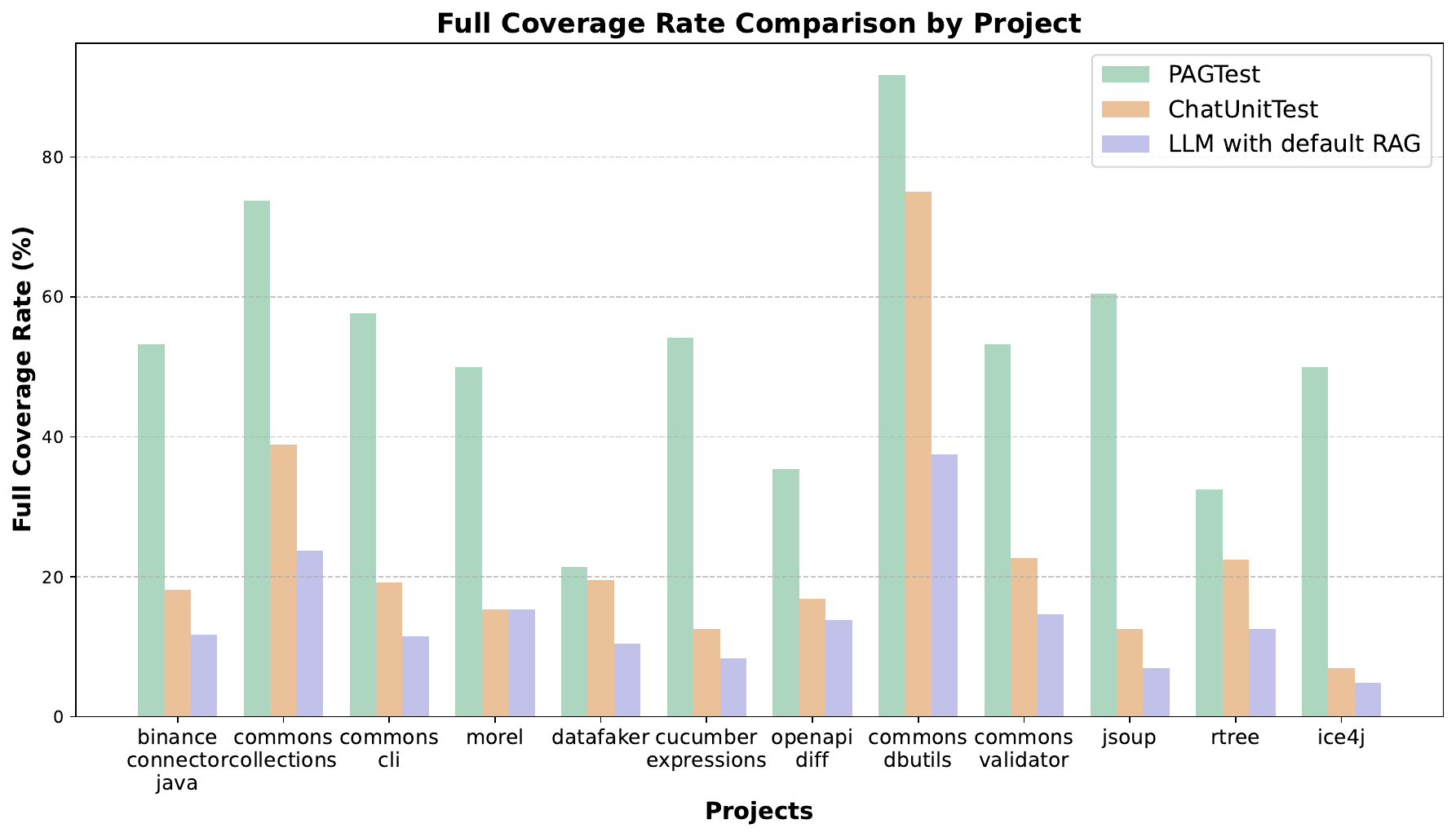}
    \vspace{-10pt}
    \caption{Project-wise Full Coverage Rate of \toolname, ChatUnitTest, and LLM with default RAG}
    \label{fig:full_coverage_comparison}
    \vspace{-10pt}
\end{figure}

Figure \ref{fig:full_coverage_comparison} compares the full coverage rates across three LLM-based tools. 
\toolname consistently achieves higher full coverage rates across nearly all projects. For example, in \texttt{binance-connector-java}, \toolname achieves 53.3\%, outperforming ChatUnitTest at 18.1\% and the LLM with default RAG at 11.7\%. Similarly, in \texttt{commons-collections}, \toolname reached 73.8\%, significantly exceeding ChatUnitTest's 38.9\%, and the LLM with default RAG's 23.8\%.





An interesting observation is that projects with abstract or interface-heavy code pose significant challenges for all tools. For instance, in the \texttt{datafaker} project, which contains numerous abstract classes and interfaces, all tools experience higher failure rates. This occurs because the tools mistakenly attempted to instantiate abstract classes, which is not permissible in Java.
In the \texttt{ice4j} project, the complexity of network protocols makes test generation particularly challenging. However, \toolname achieves a 50.0\% full coverage rate, significantly outperforming ChatUnitTest's 7.0\%. This underscores the value of incorporating existing tests when generating new tests in projects with intricate dependencies and protocols.





\subsection{Maintainability of Generated Unit Tests}
To address this research question, we evaluate the maintainability of unit tests that passed and achieved full coverage using two primary metrics: \textbf{Code Style Violations} and \textbf{Mock Density}. 

\textbf{Code Style Violations} assess how well the generated tests adhere to coding standards, 
Based on Tang et al. \cite{tang2024chatgpt}, 
these violations are categorized as follows:
\begin{itemize}[leftmargin=*]
    \item \textbf{Code Duplication}: This refers to redundant code elements, such as redefined variables, repeated imports, or duplicate logic.
    \item \textbf{Naming Convention}: This covers inconsistencies in variable, method, or class names, such as the use of non-descriptive or inconsistent names.
    \item \textbf{Miscellaneous}: This includes other style violations such as improper indentation, missing comments, or unorganized imports.
\end{itemize}
Such violations diminish the maintainability of the generated tests, making them harder to manage and less likely to be accepted by developers.


\begin{wrapfigure}{r}{0.5\textwidth}
    \centering
    \vspace{-10pt}
    \includegraphics[width=0.48\textwidth]{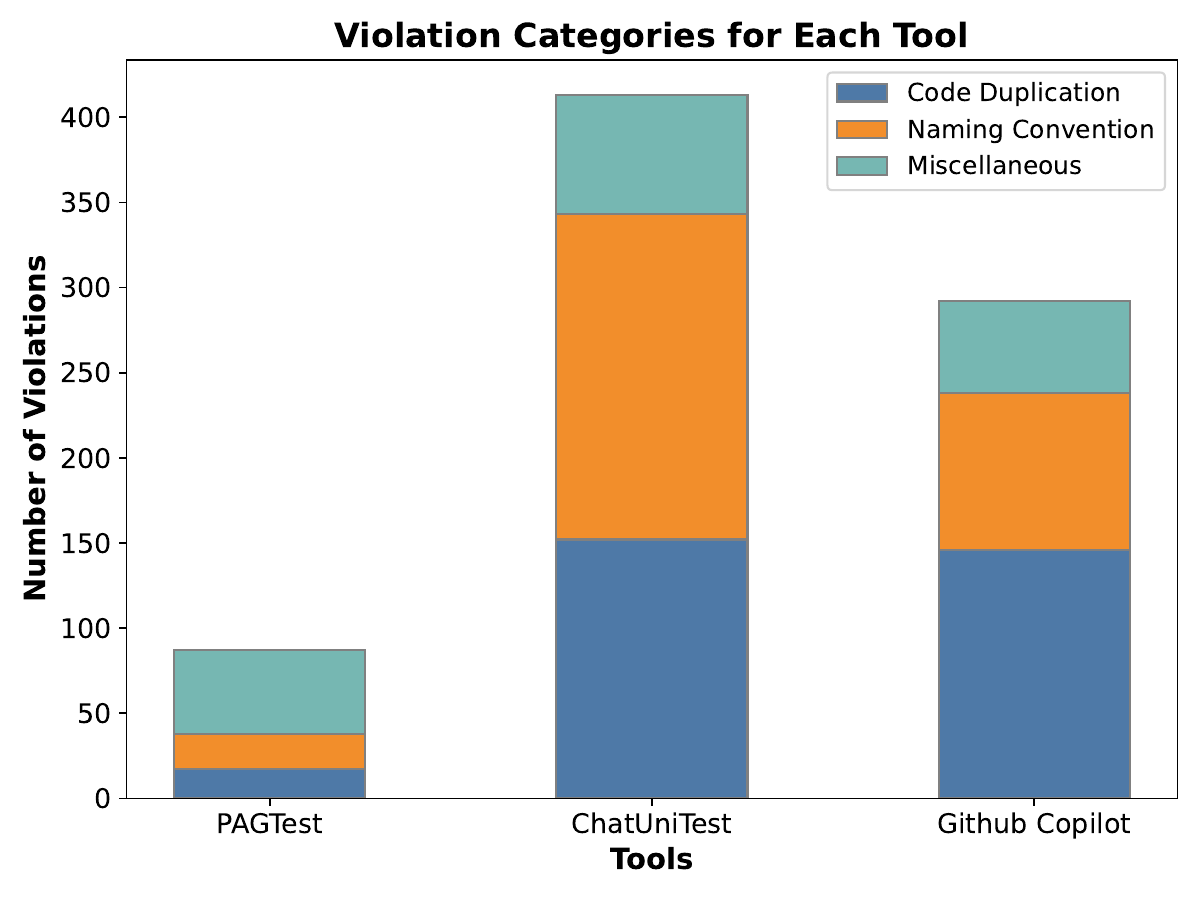}
    \vspace{-10pt}
    \caption{Code style violations across different tools.}
    \vspace{-10pt}
    \label{fig:violation_categories_chart}
\end{wrapfigure}
Following the methodology outlined by Tang et al. \cite{tang2024chatgpt}, we used CheckStyle \cite{Checkstyle} for code style checking.
We opt not to evaluate the tests produced by EvoSuite due to its strong reliance on custom auxiliary libraries and scaffolding inheritance. 
Additionally, the test cases generated by EvoSuite often suffer from non-standard naming conventions (e.g., ``test0'', ``test1''), which further detracts from their usability. Instead, we choose GitHub Copilot for our comparison.
We select 167 unit tests generated for methods that were fully covered by all three tools.
These tests are integrated back into their original test suites for evaluation.

The analysis of code violations is shown in Figure \ref{fig:violation_categories_chart}, and the detailed breakdown is as follows:
\begin{itemize}[leftmargin=*]
    \item \textbf{Code Duplication}: \toolname introduces 17 instances, while ChatUnitTest and Github Copilot result in significantly higher duplication, with 152 and 146 violations, respectively.
    \item \textbf{Naming Convention}: \toolname has 21 violations, which is substantially lower than ChatUnitTest’s 191 and Github Copilot’s 92 violations.
    \item \textbf{Miscellaneous}: \toolname again performs better, with 49 violations compared to ChatUnitTest’s 70 and Github Copilot’s 54.
\end{itemize}

These results show that \toolname generates maintainable and readable tests, with significantly fewer code style violations, especially in key features like code duplication and naming conventions.



\begin{wraptable}{r}{0.45\textwidth}
\centering
\footnotesize
\vspace{-10pt}
\caption{Mock Detection Analysis Across Tools}
\vspace{-5pt}
\label{tab:mock_analysis}
\renewcommand{\arraystretch}{1.3}
\begin{tabular}{l|c}
\hline
\textbf{Tool} & \textbf{Additional Mocks} \\ \hline
\toolname & 8 \\ \hline
ChatUniTest & 61 \\ \hline
Github Copilot & 42 \\ \hline
\end{tabular}
\vspace{-10pt}
\end{wraptable}
\textbf{Mock Density} evaluates the use of mock objects, as excessive mocking can lead to lower test readability, higher maintenance difficulty, and overly fine-grained test cases \cite{spadini2017mock}.
To further analyze the usability of the generated unit tests, we use PMD \cite{Pmd} used by Tang et al.~\cite{tang2024chatgpt} for mock detection. 
Table \ref{tab:mock_analysis} compares the additional mocks detected by each tool. 
The results clearly show that \toolname introduces only 8 additional mocks, significantly fewer than ChatUniTest, which detects 61 unnecessary mocks. Copilot, while more conservative than ChatUniTest, still adds 42 unnecessary mocks—far more than \toolname. This highlights that \toolname produces the fewest redundant mocks,
resulting in cleaner and more maintainable tests.


\subsection{Impact of Iterative Strategy}

To demonstrate the impact of the Iterative Incremental Strategy (IS), we conduct an ablation study on the projects from RQ1 (Section \ref{sec:rq1}). 

\begin{table}[htbp]
\centering
\footnotesize
\vspace{-10pt}
\caption{Comparison of Performance Metrics with and without Iterative Strategy (IS)}
\vspace{-5pt}
\resizebox{\textwidth}{!}{%
\begin{tabular}{l|c|c|c|c|c|c|c|c}
\toprule
\textbf{Project Name} & \multicolumn{2}{c|}{\textbf{Compilation and Run Error}} & \multicolumn{2}{c|}{\textbf{Assertion Error}} & \multicolumn{2}{c|}{\textbf{Successful Executions}} & \multicolumn{2}{c}{\textbf{Full Coverage}} \\ \cline{2-9}
& \textbf{Without IS} & \textbf{With IS} & \textbf{Without IS} & \textbf{With IS} & \textbf{Without IS} & \textbf{With IS} & \textbf{Without IS} & \textbf{With IS} \\ \hline
binance-connector-java & 22.6\% & 12.3\% & 35.8\% & 34.1\% & 41.6\% & 53.6\% & 38.4\% & 53.3\% \\ \hline
commons-collections & 31.7\% & 17.9\% & 6.6\% & 4.8\% & 61.7\% & 77.2\% & 60.3\% & 73.8\% \\ \hline
commons-cli & 34.6\% & 23.1\% & 11.5\% & 9.6\% & 53.9\% & 67.3\% & 50.0\% & 57.7\% \\ \hline
morel & 42.3\% & 32.7\% & 11.5\% & 9.6\% & 46.2\% & 57.7\% & 42.3\% & 50.0\% \\ \hline
datafaker & 70.8\% & 69.5\% & 9.7\% & 8.4\% & 19.5\% & 22.1\% & 18.8\% & 21.4\% \\ \hline
cucumber-expressions & 50.0\% & 37.5\% & 0.0\% & 0.0\% & 50.0\% & 62.5\% & 45.8\% & 54.2\% \\ \hline
openapi-diff & 61.5\% & 58.5\% & 7.7\% & 4.6\% & 30.8\% & 36.9\% & 29.2\% & 35.4\% \\ \hline
commons-dbutils & 4.2\% & 0.0\% & 4.2\% & 0.0\% & 91.7\% & 100.0\% & 91.7\% & 91.7\% \\ \hline
commons-validator & 32.0\% & 20.0\% & 16.0\% & 14.7\% & 52.0\% & 65.3\% & 46.7\% & 53.3\% \\ \hline
jsoup & 26.2\% & 11.3\% & 15.3\% & 12.5\% & 58.5\% & 76.2\% & 51.6\% & 60.5\% \\ \hline
rtree & 60.0\% & 57.5\% & 5.0\% & 2.5\% & 35.0\% & 40.0\% & 27.5\% & 32.5\% \\ \hline
ice4j & 50.7\% & 34.5\% & 7.0\% & 5.6\% & 42.3\% & 59.9\% & 34.5\% & 50.0\% \\ 
\bottomrule
\end{tabular}
}
\vspace{-10pt}
\label{tab:performance_metrics}
\end{table}

As shown in Table \ref{tab:performance_metrics}, IS significantly reduces the \textbf{Compilation and Run Error Rate} across most projects.
For example, in project \texttt{binance-connector-java}, with IS, the compilation and run error reduces from 22.6\% to 12.3\%, and \texttt{jsoup} drops from 26.2\% to 11.3\%. Similarly, showing that IS helps in generating more syntactically and semantically correct tests. The \textbf{Assertion Error Rate} is also slightly decreased, with \texttt{commons-validator} reducing from 16.0\% to 14.7\% and \texttt{rtree} from 5.0\% to 2.5\%, indicating a reduction in logic errors within test assertions.
For the \textbf{Successful Execution Rate}, IS demonstrates notable improvements. Projects like \texttt{commons-collections} and \texttt{jsoup} experience increases from 61.7\% to 77.2\% and from 58.5\% to 76.2\%, respectively. This shows that IS enhances the ability to generate tests that successfully execute, covering more code paths without encountering runtime issues. The \textbf{Full Coverage Rate} also reflects positive impact, with \texttt{binance-connector-java} improving from 38.4\% to 53.3\% and \texttt{commons-cli} from 50.0\% to 57.7\%. These results confirm that IS not only helps generate executable tests but also boosts their ability to provide full coverage of the tested code.

Overall, the results confirm that the Iterative Incremental Strategy improves test quality by reducing errors and enhancing both coverage success and overall coverage.

\subsection{Exploring Patterns in Property Relationships}


To understand the reason why \toolname is effective, we further analyze the patterns in \emph{property relationships}.
To do so, we manually analyze the paired methods with property relations and summarize the property patterns.

\subsubsection{Study Design}


To minimize bias and ensure reliability, we employ several rigorous strategies in our study design, referencing methodologies from prior work \cite{ran2024guardian, xu2024mr}:

\begin{itemize}[leftmargin=*]
    \item \textbf{Participant Selection and Training}:
        A team of 5 graduate and doctoral researchers, all majoring in software engineering with relevant expertise, are selected as participants. Comprehensive training sessions are conducted, covering the definition of property relationships, the review process, and the criteria for validation. Example cases were used to illustrate expected outcomes.
    
    \item \textbf{Cross-Validation}:
            Each method relationship generated by the LLM is reviewed independently by two participants. In cases where their assessments differed, a third researcher is involved in resolving the disagreement through discussion. This ensured a high level of agreement. 
    
    \item \textbf{Random Sampling and Multiple Rounds}:
           We randomly select representative 300 methods.
           The analysis is conducted in two rounds, with different participants reviewing the same relationships in the second round to verify consistency and identify any discrepancies.

\end{itemize}

\subsubsection{Analysis and Implications}

As shown in Table \ref{tab:property_patterns}, we select the top 6 patterns in property relationships that demonstrate how test logic can be enhanced across different phases. The first three patterns, Structural Similarity, Behavioral Similarity, and Substitutability, allow for enhancements across all three test phases: \textit{Given}, \textit{When}, and \textit{Then}. These patterns highlight the potential for reusing test logic comprehensively. In contrast, the last three patterns, Exception Handling Similarity, Resource Access Similarity, and Dependency, enhance specific phases of testing. These targeted enhancements focus on key areas like exception handling and resource management, ensuring that testing remains both efficient and contextually appropriate for each method's unique characteristics.

\begin{table}[htbp]
\centering
\footnotesize
\vspace{-10pt}
\caption{Patterns in Property Relationships}
\vspace{-6pt}
\begin{tabularx}{\textwidth}{|p{1.5cm}|X|X|p{1cm}|p{0.7cm}|}
\hline


\textbf{Pattern}
& \textbf{Description} & \textbf{Examples} & \textbf{Phases} & \textbf{Count} \\ \hline

\textbf{Structural Similarity} & Methods share internal structures or control flows, even if operations differ. & \textit{readFile()} vs. \textit{writeFile()}: Both handle files with similar control structures but perform opposite tasks. & Given, When, Then & 320 \\ \hline

\textbf{Behavioral Similarity} & Methods achieve similar outcomes but operate on different data or contexts. & \textit{fetchUserName()} vs. \textit{fetchUserEmail()}: Both retrieve user information but return different data types. & Given, When, Then & 617 \\ \hline

\textbf{Substituta- bility} & Methods serve the same purpose with different implementations. & \textit{mergeSort()} vs. \textit{quickSort()}: Both sort data but use different algorithms. & Given, When, Then & 124 \\ \hline

\textbf{Exception Handling Similarity} & Methods handle exceptions in similar ways but under different contexts. & \textit{connectToServer()} vs. \textit{downloadFile()}: Both handle {TimeoutException}, though one is for establishing a server connection and the other is for downloading a file. & Then & 357 \\ \hline

\textbf{Resource Access Similarity} & Methods access the same resource but perform entirely different operations on it. & \textit{lockFile()} vs. \textit{deleteFile()}: Both methods access a file, but one locks it for editing, while the other deletes it. & Given, When & 400 \\ \hline

\textbf{Dependency} & Methods rely on similar initialization or preconditions for operation. & \textit{initializeIterator()} vs. \textit{asMultipleUseIterable()}: Both require iterator initialization. & Given & 351 \\ \hline

\end{tabularx}
\vspace{-10pt}
\label{tab:property_patterns}
\end{table}

\section{DISCUSSION}

\subsection{Threats to Validity}
The threats to the validity of \toolname primarily involve its generalization to other LLMs and the diversity of the projects used in the evaluation. Due to budget constraints, we utilize DeepSeek V2.5, which offers a balance between performance and cost. However, using different LLMs may yield varying results and could impact the overall effectiveness of the tool.
Additionally, the range of projects may introduce bias, as they vary in complexity and existing unit test coverage. The presence of existing unit tests can affect \toolname's performance, as it relies on these tests to establish property relationships and guide the generation process. Future work will involve expanding the dataset and testing with other LLMs to better assess the generalizability of \toolname.
\subsection{Extensibility}

\paragraph{Function-Level Enhancement} 
While our paper demonstrates class-level and repository-level enhancement relationships, for a focal method, its own existing test cases can also provide valuable insights for generating new tests. Although this does not depend on the defined \textit{property relationships}, it leverages task-specific context effectively.

\paragraph{Extend to Other Code-Related Tasks} 
At a higher level of abstraction, our approach extends beyond unit test generation. By defining \textit{property relationships} between methods within code-specific tasks, the framework is applicable to other code-related tasks, not limited to test generation.



\section{RELATED WORK}
\subsection{Unit Test Generation}
Unit testing focuses on testing individual hardware or software units, or groups of related units \cite{olan2003unit}. Several test case generation approaches have been proposed, including random-based methods \cite{davis2023nanofuzz, wei2022free} and constraint-driven techniques \cite{hwang2021justgen}. Approaches such as DART \cite{godefroid2005dart} and KLEE \cite{ma2015grt}, which leverage symbolic execution, often face challenges like path explosion \cite{baldoni2018survey}. Search-based software testing (SBST) techniques \cite{chen2023compiler, feldmeier2022neuroevolution, lemieux2023codamosa, lin2023route, sun2023evolutionary, yandrapally2022fragment, zhou2022selectively}, such as EvoSuite \cite{andrews2011genetic} and evolutionary testing \cite{tonella2004evolutionary}, mitigate some of these issues but struggle with broad search spaces and high computational costs \cite{mcminn2011search}. In contrast, deep learning-based methods \cite{blasi2022call, feldmeier2022neuroevolution, wang2020automatic, ye2023generative, zhao2022avgust}, like AthenaTest \cite{tufano2020unit}, leverage neural models to generate diverse test inputs and better capture functional program intent.

Recently, large language model (LLM)-based techniques have gained popularity in testing \cite{guilherme2023initial, deng2023large}. Xie et al. \cite{xie2023chatunitest} introduced ChatUniTest, which uses ChatGPT to generate Java-compliant test cases after static analysis. Yuan et al. \cite{yuan2023no} proposed ChatTester, which refines initial test cases with ChatGPT, sometimes outperforming traditional SBST methods. Hybrid approaches like TELPA \cite{yang2024enhancing}, uses program analysis with LLM by integrating refined coun-
terexamples into prompts, guiding the LLM to generate diverse tests for hard-to-cover branches, and HITS \cite{wang2024hits} decomposes focal methods into slices and ask the LLM to generate tests slice by
slice, improving coverage for complex methods. While these tools have demonstrated promising results in specific scenarios, they often struggle with deeply understanding the focal method and fail to account for existing test cases. This leads to generated tests lacking practical usability and alignment with existing testing infrastructure. In contrast, our tool not only comprehends the focal method in-depth, but also leverages existing test cases to increase the correctness and practical applicability of the generated tests.

\subsection{Retrieval-Augmented Generation (RAG) for Code Tasks}
RAG systems, particularly Retrieval-Augmented Code Generation (RACG), enhance large language models (LLMs) by retrieving relevant code snippets or structures from repositories \cite{jiang2024survey}. Existing approaches include similarity-based methods \cite{robertson2009probabilistic, guo2022unixcoder}, such as using the BM25 algorithm to perform similarity retrieval \cite{jimenez2023swe}, as well as vector similarity-based retrieval \cite{pan2024enhancing}. Additionally, AST-based retrieval approaches, such as AutoCodeRover \cite{zhang2024autocoderover}, leverage AST structures to retrieve relevant code contexts. Some methods design specific tools to perform more targeted retrieval, as in the case of MASAI \cite{arora2024masai}, while CODEXGRAPH \cite{liu2024codexgraph} uses code graph databases to allow for flexible and powerful retrieval to get more code structure information. Further approaches involve retrieving based on ``repo-specific semantic graphs'' \cite{liang2024repofuserepositorylevelcodecompletion}.
Our tool extends traditional retrieval methods by considering task-specific contexts of code and defining property relationships to guide the retrieval process, offering a more tailored approach to enhancing test generation.

\section{CONCLUSION}

We presented \textit{Property-Based Retrieval Augmentation for Unit Test Generation}, a novel approach that enhances the correctness, completeness, and maintainability of unit tests by leveraging property relationships between methods. By extending LLM-based Retrieval-Augmented Generation (RAG) with code-specific relationships such as behavioral similarities and structural dependencies, our method improved test generation through the \textit{Given}, \textit{When}, and \textit{Then} phases. \toolname further employed an iterative strategy, using newly generated tests to guide future ones. Our evaluation across 12 open-source projects demonstrated that \toolname significantly outperformed existing tools, offering a promising direction for more reliable and context-aware test generation.

Moreover, our work introduced a novel code-context-aware retrieval mechanism for LLMs, offering valuable insights and potential applications for other code-related tasks.



\bibliographystyle{ACM-Reference-Format}
\bibliography{sample-base}


\begin{thebibliography}{61}


\ifx \showCODEN    \undefined \def \showCODEN     #1{\unskip}     \fi
\ifx \showDOI      \undefined \def \showDOI       #1{#1}\fi
\ifx \showISBNx    \undefined \def \showISBNx     #1{\unskip}     \fi
\ifx \showISBNxiii \undefined \def \showISBNxiii  #1{\unskip}     \fi
\ifx \showISSN     \undefined \def \showISSN      #1{\unskip}     \fi
\ifx \showLCCN     \undefined \def \showLCCN      #1{\unskip}     \fi
\ifx \shownote     \undefined \def \shownote      #1{#1}          \fi
\ifx \showarticletitle \undefined \def \showarticletitle #1{#1}   \fi
\ifx \showURL      \undefined \def \showURL       {\relax}        \fi
\providecommand\bibfield[2]{#2}
\providecommand\bibinfo[2]{#2}
\providecommand\natexlab[1]{#1}
\providecommand\showeprint[2][]{arXiv:#2}

\bibitem[Andrews et~al\mbox{.}(2011)]%
        {andrews2011genetic}
\bibfield{author}{\bibinfo{person}{J.~H. Andrews}, \bibinfo{person}{T. Menzies}, {and} \bibinfo{person}{F.~C.~H. Li}.} \bibinfo{year}{2011}\natexlab{}.
\newblock \showarticletitle{Genetic algorithms for randomized unit testing}.
\newblock \bibinfo{journal}{\emph{IEEE Transactions on Software Engineering}} \bibinfo{volume}{37}, \bibinfo{number}{1} (\bibinfo{year}{2011}), \bibinfo{pages}{80--94}.
\newblock


\bibitem[Arora et~al\mbox{.}(2024)]%
        {arora2024masai}
\bibfield{author}{\bibinfo{person}{Daman Arora}, \bibinfo{person}{Atharv Sonwane}, \bibinfo{person}{Nalin Wadhwa}, \bibinfo{person}{Abhav Mehrotra}, \bibinfo{person}{Saiteja Utpala}, \bibinfo{person}{Ramakrishna Bairi}, \bibinfo{person}{Aditya Kanade}, {and} \bibinfo{person}{Nagarajan Natarajan}.} \bibinfo{year}{2024}\natexlab{}.
\newblock \showarticletitle{MASAI: Modular Architecture for Software-engineering AI Agents}.
\newblock \bibinfo{journal}{\emph{arXiv preprint arXiv:2406.11638}} (\bibinfo{year}{2024}).
\newblock


\bibitem[Baldoni et~al\mbox{.}(2018)]%
        {baldoni2018survey}
\bibfield{author}{\bibinfo{person}{Riccardo Baldoni}, \bibinfo{person}{Enrico Coppa}, \bibinfo{person}{Domenico~C D’elia}, {et~al\mbox{.}}} \bibinfo{year}{2018}\natexlab{}.
\newblock \showarticletitle{A survey of symbolic execution techniques}.
\newblock \bibinfo{journal}{\emph{ACM Computing Surveys (CSUR)}} \bibinfo{volume}{51}, \bibinfo{number}{3} (\bibinfo{year}{2018}), \bibinfo{pages}{1--39}.
\newblock


\bibitem[Blasi et~al\mbox{.}(2022)]%
        {blasi2022call}
\bibfield{author}{\bibinfo{person}{Arianna Blasi}, \bibinfo{person}{Alessandra Gorla}, \bibinfo{person}{Michael~D Ernst}, {and} \bibinfo{person}{Mauro Pezz{\`e}}.} \bibinfo{year}{2022}\natexlab{}.
\newblock \showarticletitle{Call me maybe: Using nlp to automatically generate unit test cases respecting temporal constraints}. In \bibinfo{booktitle}{\emph{Proceedings of the 37th IEEE/ACM International Conference on Automated Software Engineering}}. \bibinfo{pages}{1--11}.
\newblock


\bibitem[ChatBotArena(2024)]%
        {ChatBotArena}
\bibfield{author}{\bibinfo{person}{ChatBotArena}.} \bibinfo{year}{2024}\natexlab{}.
\newblock \bibinfo{title}{"Chatbot Arena LLM Leaderboard: Community-driven Evaluation for Best LLM and AI chatbots"}.
\newblock
\newblock
\urldef\tempurl%
\url{https://lmarena.ai/}
\showURL{%
\tempurl}
\newblock
\shownote{Accessed: 2024}.


\bibitem[Checkstyle(1999)]%
        {Checkstyle}
\bibfield{author}{\bibinfo{person}{Checkstyle}.} \bibinfo{year}{1999}\natexlab{}.
\newblock \bibinfo{title}{"Code conventions for the java programming language"}.
\newblock
\newblock
\urldef\tempurl%
\url{https://checkstyle.sourceforge.io/}
\showURL{%
\tempurl}
\newblock
\shownote{Accessed: 2024}.


\bibitem[Chen et~al\mbox{.}(2023a)]%
        {chen2023you}
\bibfield{author}{\bibinfo{person}{Jiuhai Chen}, \bibinfo{person}{Lichang Chen}, \bibinfo{person}{Heng Huang}, {and} \bibinfo{person}{Tianyi Zhou}.} \bibinfo{year}{2023}\natexlab{a}.
\newblock \showarticletitle{When do you need chain-of-thought prompting for chatgpt?}
\newblock \bibinfo{journal}{\emph{arXiv preprint arXiv:2304.03262}} (\bibinfo{year}{2023}).
\newblock


\bibitem[Chen et~al\mbox{.}(2023b)]%
        {chen2023compiler}
\bibfield{author}{\bibinfo{person}{Junjie Chen}, \bibinfo{person}{Chenyao Suo}, \bibinfo{person}{Jiajun Jiang}, \bibinfo{person}{Peiqi Chen}, {and} \bibinfo{person}{Xingjian Li}.} \bibinfo{year}{2023}\natexlab{b}.
\newblock \showarticletitle{Compiler test-program generation via memoized configuration search}. In \bibinfo{booktitle}{\emph{2023 IEEE/ACM 45th International Conference on Software Engineering (ICSE)}}. IEEE, \bibinfo{pages}{2035--2047}.
\newblock


\bibitem[Davis et~al\mbox{.}(2023)]%
        {davis2023nanofuzz}
\bibfield{author}{\bibinfo{person}{Matthew~C Davis}, \bibinfo{person}{Sangheon Choi}, \bibinfo{person}{Sam Estep}, \bibinfo{person}{Brad~A Myers}, {and} \bibinfo{person}{Joshua Sunshine}.} \bibinfo{year}{2023}\natexlab{}.
\newblock \showarticletitle{NaNofuzz: A Usable Tool for Automatic Test Generation}. In \bibinfo{booktitle}{\emph{Proceedings of the 31st ACM Joint European Software Engineering Conference and Symposium on the Foundations of Software Engineering}}. \bibinfo{pages}{1114--1126}.
\newblock


\bibitem[Deng et~al\mbox{.}(2023)]%
        {deng2023large}
\bibfield{author}{\bibinfo{person}{Yinlin Deng}, \bibinfo{person}{Chunqiu~Steven Xia}, \bibinfo{person}{Haoran Peng}, \bibinfo{person}{Chenyuan Yang}, {and} \bibinfo{person}{Lingming Zhang}.} \bibinfo{year}{2023}\natexlab{}.
\newblock \showarticletitle{Large language models are zero-shot fuzzers: Fuzzing deep-learning libraries via large language models}. In \bibinfo{booktitle}{\emph{Proceedings of the 32nd ACM SIGSOFT international symposium on software testing and analysis}}. \bibinfo{pages}{423--435}.
\newblock


\bibitem[Feldmeier and Fraser(2022)]%
        {feldmeier2022neuroevolution}
\bibfield{author}{\bibinfo{person}{Patric Feldmeier} {and} \bibinfo{person}{Gordon Fraser}.} \bibinfo{year}{2022}\natexlab{}.
\newblock \showarticletitle{Neuroevolution-based generation of tests and oracles for games}. In \bibinfo{booktitle}{\emph{Proceedings of the 37th IEEE/ACM International Conference on Automated Software Engineering}}. \bibinfo{pages}{1--13}.
\newblock


\bibitem[Fraser and Arcuri(2011)]%
        {fraser2011evosuite}
\bibfield{author}{\bibinfo{person}{Gordon Fraser} {and} \bibinfo{person}{Andrea Arcuri}.} \bibinfo{year}{2011}\natexlab{}.
\newblock \showarticletitle{Evosuite: automatic test suite generation for object-oriented software}. In \bibinfo{booktitle}{\emph{Proceedings of the 19th ACM SIGSOFT symposium and the 13th European conference on Foundations of software engineering}}. \bibinfo{pages}{416--419}.
\newblock


\bibitem[Github(2024)]%
        {GithubCopiot}
\bibfield{author}{\bibinfo{person}{Github}.} \bibinfo{year}{2024}\natexlab{}.
\newblock \bibinfo{title}{"The world’s most widely adopted AI developer tool."}.
\newblock
\newblock
\urldef\tempurl%
\url{https://github.com/features/copilot}
\showURL{%
\tempurl}
\newblock
\shownote{Accessed: 2024}.


\bibitem[Godefroid et~al\mbox{.}(2005)]%
        {godefroid2005dart}
\bibfield{author}{\bibinfo{person}{Patrice Godefroid}, \bibinfo{person}{Nils Klarlund}, {and} \bibinfo{person}{Koushik Sen}.} \bibinfo{year}{2005}\natexlab{}.
\newblock \showarticletitle{{DART: Directed automated random testing}}. In \bibinfo{booktitle}{\emph{Proceedings of the 2005 ACM SIGPLAN conference on Programming language design and implementation}}. ACM, \bibinfo{pages}{213--223}.
\newblock


\bibitem[Grano et~al\mbox{.}(2018)]%
        {grano2018empirical}
\bibfield{author}{\bibinfo{person}{Giovanni Grano}, \bibinfo{person}{Simone Scalabrino}, \bibinfo{person}{Harald~C Gall}, {and} \bibinfo{person}{Rocco Oliveto}.} \bibinfo{year}{2018}\natexlab{}.
\newblock \showarticletitle{An empirical investigation on the readability of manual and generated test cases}. In \bibinfo{booktitle}{\emph{Proceedings of the 26th Conference on Program Comprehension}}. \bibinfo{pages}{348--351}.
\newblock


\bibitem[Gu et~al\mbox{.}(2024)]%
        {gu2024testart}
\bibfield{author}{\bibinfo{person}{Siqi Gu}, \bibinfo{person}{Chunrong Fang}, \bibinfo{person}{Quanjun Zhang}, \bibinfo{person}{Fangyuan Tian}, {and} \bibinfo{person}{Zhenyu Chen}.} \bibinfo{year}{2024}\natexlab{}.
\newblock \showarticletitle{TestART: Improving LLM-based Unit Test via Co-evolution of Automated Generation and Repair Iteration}.
\newblock \bibinfo{journal}{\emph{arXiv preprint arXiv:2408.03095}} (\bibinfo{year}{2024}).
\newblock


\bibitem[Guilherme and Vincenzi(2023)]%
        {guilherme2023initial}
\bibfield{author}{\bibinfo{person}{Vitor Guilherme} {and} \bibinfo{person}{Auri Vincenzi}.} \bibinfo{year}{2023}\natexlab{}.
\newblock \showarticletitle{An initial investigation of ChatGPT unit test generation capability}. In \bibinfo{booktitle}{\emph{Proceedings of the 8th Brazilian Symposium on Systematic and Automated Software Testing}}. \bibinfo{pages}{15--24}.
\newblock


\bibitem[Guo et~al\mbox{.}(2022)]%
        {guo2022unixcoder}
\bibfield{author}{\bibinfo{person}{Daya Guo}, \bibinfo{person}{Shuai Lu}, \bibinfo{person}{Nan Duan}, \bibinfo{person}{Yanlin Wang}, \bibinfo{person}{Ming Zhou}, {and} \bibinfo{person}{Jian Yin}.} \bibinfo{year}{2022}\natexlab{}.
\newblock \showarticletitle{Unixcoder: Unified cross-modal pre-training for code representation}.
\newblock \bibinfo{journal}{\emph{arXiv preprint arXiv:2203.03850}} (\bibinfo{year}{2022}).
\newblock


\bibitem[Hwang et~al\mbox{.}(2021)]%
        {hwang2021justgen}
\bibfield{author}{\bibinfo{person}{Sungjae Hwang}, \bibinfo{person}{Sungho Lee}, \bibinfo{person}{Jihoon Kim}, {and} \bibinfo{person}{Sukyoung Ryu}.} \bibinfo{year}{2021}\natexlab{}.
\newblock \showarticletitle{Justgen: Effective test generation for unspecified JNI behaviors on jvms}. In \bibinfo{booktitle}{\emph{2021 IEEE/ACM 43rd International Conference on Software Engineering (ICSE)}}. IEEE, \bibinfo{pages}{1708--1718}.
\newblock


\bibitem[jedi(2024)]%
        {jedi}
\bibfield{author}{\bibinfo{person}{jedi}.} \bibinfo{year}{2024}\natexlab{}.
\newblock \bibinfo{title}{"Awesome autocompletion, static analysis and refactoring library for python"}.
\newblock
\newblock
\urldef\tempurl%
\url{https://github.com/davidhalter/jedi}
\showURL{%
\tempurl}
\newblock
\shownote{Accessed: 2024}.


\bibitem[Jiang et~al\mbox{.}(2024)]%
        {jiang2024survey}
\bibfield{author}{\bibinfo{person}{Juyong Jiang}, \bibinfo{person}{Fan Wang}, \bibinfo{person}{Jiasi Shen}, \bibinfo{person}{Sungju Kim}, {and} \bibinfo{person}{Sunghun Kim}.} \bibinfo{year}{2024}\natexlab{}.
\newblock \showarticletitle{A Survey on Large Language Models for Code Generation}.
\newblock \bibinfo{journal}{\emph{arXiv preprint arXiv:2406.00515}} (\bibinfo{year}{2024}).
\newblock


\bibitem[Jimenez et~al\mbox{.}(2023)]%
        {jimenez2023swe}
\bibfield{author}{\bibinfo{person}{Carlos~E Jimenez}, \bibinfo{person}{John Yang}, \bibinfo{person}{Alexander Wettig}, \bibinfo{person}{Shunyu Yao}, \bibinfo{person}{Kexin Pei}, \bibinfo{person}{Ofir Press}, {and} \bibinfo{person}{Karthik Narasimhan}.} \bibinfo{year}{2023}\natexlab{}.
\newblock \showarticletitle{Swe-bench: Can language models resolve real-world github issues?}
\newblock \bibinfo{journal}{\emph{arXiv preprint arXiv:2310.06770}} (\bibinfo{year}{2023}).
\newblock


\bibitem[Lemieux et~al\mbox{.}(2023)]%
        {lemieux2023codamosa}
\bibfield{author}{\bibinfo{person}{Caroline Lemieux}, \bibinfo{person}{Jeevana~Priya Inala}, \bibinfo{person}{Shuvendu~K Lahiri}, {and} \bibinfo{person}{Siddhartha Sen}.} \bibinfo{year}{2023}\natexlab{}.
\newblock \showarticletitle{Codamosa: Escaping coverage plateaus in test generation with pre-trained large language models}. In \bibinfo{booktitle}{\emph{2023 IEEE/ACM 45th International Conference on Software Engineering (ICSE)}}. IEEE, \bibinfo{pages}{919--931}.
\newblock


\bibitem[Liang et~al\mbox{.}(2024)]%
        {liang2024repofuserepositorylevelcodecompletion}
\bibfield{author}{\bibinfo{person}{Ming Liang}, \bibinfo{person}{Xiaoheng Xie}, \bibinfo{person}{Gehao Zhang}, \bibinfo{person}{Xunjin Zheng}, \bibinfo{person}{Peng Di}, \bibinfo{person}{wei jiang}, \bibinfo{person}{Hongwei Chen}, \bibinfo{person}{Chengpeng Wang}, {and} \bibinfo{person}{Gang Fan}.} \bibinfo{year}{2024}\natexlab{}.
\newblock \bibinfo{title}{RepoFuse: Repository-Level Code Completion with Fused Dual Context}.
\newblock
\newblock
\showeprint[arxiv]{2402.14323}~[cs.SE]
\urldef\tempurl%
\url{https://arxiv.org/abs/2402.14323}
\showURL{%
\tempurl}


\bibitem[Lin et~al\mbox{.}(2023)]%
        {lin2023route}
\bibfield{author}{\bibinfo{person}{Jun-Wei Lin}, \bibinfo{person}{Navid Salehnamadi}, {and} \bibinfo{person}{Sam Malek}.} \bibinfo{year}{2023}\natexlab{}.
\newblock \showarticletitle{Route: Roads not taken in ui testing}.
\newblock \bibinfo{journal}{\emph{ACM Transactions on Software Engineering and Methodology}} \bibinfo{volume}{32}, \bibinfo{number}{3} (\bibinfo{year}{2023}), \bibinfo{pages}{1--25}.
\newblock


\bibitem[Liu et~al\mbox{.}(2024b)]%
        {liu2024llm}
\bibfield{author}{\bibinfo{person}{Kaibo Liu}, \bibinfo{person}{Yiyang Liu}, \bibinfo{person}{Zhenpeng Chen}, \bibinfo{person}{Jie~M Zhang}, \bibinfo{person}{Yudong Han}, \bibinfo{person}{Yun Ma}, \bibinfo{person}{Ge Li}, {and} \bibinfo{person}{Gang Huang}.} \bibinfo{year}{2024}\natexlab{b}.
\newblock \showarticletitle{LLM-Powered Test Case Generation for Detecting Tricky Bugs}.
\newblock \bibinfo{journal}{\emph{arXiv preprint arXiv:2404.10304}} (\bibinfo{year}{2024}).
\newblock


\bibitem[Liu et~al\mbox{.}(2024a)]%
        {liu2024codexgraph}
\bibfield{author}{\bibinfo{person}{Xiangyan Liu}, \bibinfo{person}{Bo Lan}, \bibinfo{person}{Zhiyuan Hu}, \bibinfo{person}{Yang Liu}, \bibinfo{person}{Zhicheng Zhang}, \bibinfo{person}{Wenmeng Zhou}, \bibinfo{person}{Fei Wang}, {and} \bibinfo{person}{Michael Shieh}.} \bibinfo{year}{2024}\natexlab{a}.
\newblock \showarticletitle{CodexGraph: Bridging Large Language Models and Code Repositories via Code Graph Databases}.
\newblock \bibinfo{journal}{\emph{arXiv preprint arXiv:2408.03910}} (\bibinfo{year}{2024}).
\newblock


\bibitem[Lukasczyk and Fraser(2022)]%
        {lukasczyk2022pynguin}
\bibfield{author}{\bibinfo{person}{Stephan Lukasczyk} {and} \bibinfo{person}{Gordon Fraser}.} \bibinfo{year}{2022}\natexlab{}.
\newblock \showarticletitle{Pynguin: Automated unit test generation for python}. In \bibinfo{booktitle}{\emph{Proceedings of the ACM/IEEE 44th International Conference on Software Engineering: Companion Proceedings}}. \bibinfo{pages}{168--172}.
\newblock


\bibitem[Ma et~al\mbox{.}(2015)]%
        {ma2015grt}
\bibfield{author}{\bibinfo{person}{Lin Ma}, \bibinfo{person}{Cyril Artho}, \bibinfo{person}{Chao Zhang}, {et~al\mbox{.}}} \bibinfo{year}{2015}\natexlab{}.
\newblock \showarticletitle{{Grt: Program-analysis-guided random testing}}. In \bibinfo{booktitle}{\emph{2015 30th IEEE/ACM International Conference on Automated Software Engineering (ASE)}}. IEEE, \bibinfo{pages}{212--223}.
\newblock


\bibitem[McMinn(2011)]%
        {mcminn2011search}
\bibfield{author}{\bibinfo{person}{Phil McMinn}.} \bibinfo{year}{2011}\natexlab{}.
\newblock \showarticletitle{Search-based software testing: Past, present and future}. In \bibinfo{booktitle}{\emph{2011 IEEE Fourth International Conference on Software Testing, Verification and Validation Workshops}}. \bibinfo{publisher}{IEEE}, \bibinfo{pages}{153--163}.
\newblock


\bibitem[microsoft(2024)]%
        {LSP}
\bibfield{author}{\bibinfo{person}{microsoft}.} \bibinfo{year}{2024}\natexlab{}.
\newblock \bibinfo{title}{"Language Server Protocol."}.
\newblock
\newblock
\urldef\tempurl%
\url{https://microsoft.github.io/language-server-protocol}
\showURL{%
\tempurl}
\newblock
\shownote{Accessed: 2024}.


\bibitem[N{\'e}ron et~al\mbox{.}(2015)]%
        {neron2015theory}
\bibfield{author}{\bibinfo{person}{Pierre N{\'e}ron}, \bibinfo{person}{Andrew Tolmach}, \bibinfo{person}{Eelco Visser}, {and} \bibinfo{person}{Guido Wachsmuth}.} \bibinfo{year}{2015}\natexlab{}.
\newblock \showarticletitle{A theory of name resolution}. In \bibinfo{booktitle}{\emph{Programming Languages and Systems: 24th European Symposium on Programming, ESOP 2015, Held as Part of the European Joint Conferences on Theory and Practice of Software, ETAPS 2015, London, UK, April 11-18, 2015, Proceedings 24}}. Springer, \bibinfo{pages}{205--231}.
\newblock


\bibitem[Olan(2003)]%
        {olan2003unit}
\bibfield{author}{\bibinfo{person}{Michael Olan}.} \bibinfo{year}{2003}\natexlab{}.
\newblock \showarticletitle{Unit testing: test early, test often}.
\newblock \bibinfo{journal}{\emph{Journal of Computing Sciences in Colleges}} \bibinfo{volume}{19}, \bibinfo{number}{2} (\bibinfo{year}{2003}), \bibinfo{pages}{319--328}.
\newblock


\bibitem[Pacheco and Ernst(2007)]%
        {pacheco2007randoop}
\bibfield{author}{\bibinfo{person}{Carlos Pacheco} {and} \bibinfo{person}{Michael~D Ernst}.} \bibinfo{year}{2007}\natexlab{}.
\newblock \showarticletitle{Randoop: feedback-directed random testing for Java}. In \bibinfo{booktitle}{\emph{Companion to the 22nd ACM SIGPLAN conference on Object-oriented programming systems and applications companion}}. \bibinfo{pages}{815--816}.
\newblock


\bibitem[Pan et~al\mbox{.}(2024)]%
        {pan2024enhancing}
\bibfield{author}{\bibinfo{person}{Zhiyuan Pan}, \bibinfo{person}{Xing Hu}, \bibinfo{person}{Xin Xia}, {and} \bibinfo{person}{Xiaohu Yang}.} \bibinfo{year}{2024}\natexlab{}.
\newblock \showarticletitle{Enhancing Repository-Level Code Generation with Integrated Contextual Information}.
\newblock \bibinfo{journal}{\emph{arXiv preprint arXiv:2406.03283}} (\bibinfo{year}{2024}).
\newblock


\bibitem[Pmd(2023)]%
        {Pmd}
\bibfield{author}{\bibinfo{person}{Pmd}.} \bibinfo{year}{2023}\natexlab{}.
\newblock \bibinfo{title}{"An extensible cross-language static code analyzer."}.
\newblock
\newblock
\urldef\tempurl%
\url{https://pmd.github.io/}
\showURL{%
\tempurl}
\newblock
\shownote{Accessed: 2023}.


\bibitem[Ran et~al\mbox{.}(2024)]%
        {ran2024guardian}
\bibfield{author}{\bibinfo{person}{Dezhi Ran}, \bibinfo{person}{Hao Wang}, \bibinfo{person}{Zihe Song}, \bibinfo{person}{Mengzhou Wu}, \bibinfo{person}{Yuan Cao}, \bibinfo{person}{Ying Zhang}, \bibinfo{person}{Wei Yang}, {and} \bibinfo{person}{Tao Xie}.} \bibinfo{year}{2024}\natexlab{}.
\newblock \showarticletitle{Guardian: A Runtime Framework for LLM-Based UI Exploration}. In \bibinfo{booktitle}{\emph{Proceedings of the 33rd ACM SIGSOFT International Symposium on Software Testing and Analysis}}. \bibinfo{pages}{958--970}.
\newblock


\bibitem[redisson(2024)]%
        {Redisson}
\bibfield{author}{\bibinfo{person}{redisson}.} \bibinfo{year}{2024}\natexlab{}.
\newblock \bibinfo{title}{"Easy Redis Java client and Real-Time Data Platform"}.
\newblock
\newblock
\urldef\tempurl%
\url{https://github.com/redisson/redisson}
\showURL{%
\tempurl}
\newblock
\shownote{Accessed: 2024}.


\bibitem[Robertson et~al\mbox{.}(2009)]%
        {robertson2009probabilistic}
\bibfield{author}{\bibinfo{person}{Stephen Robertson}, \bibinfo{person}{Hugo Zaragoza}, {et~al\mbox{.}}} \bibinfo{year}{2009}\natexlab{}.
\newblock \showarticletitle{The probabilistic relevance framework: BM25 and beyond}.
\newblock \bibinfo{journal}{\emph{Foundations and Trends{\textregistered} in Information Retrieval}} \bibinfo{volume}{3}, \bibinfo{number}{4} (\bibinfo{year}{2009}), \bibinfo{pages}{333--389}.
\newblock


\bibitem[Servantez et~al\mbox{.}(2024)]%
        {servantez2024chain}
\bibfield{author}{\bibinfo{person}{Sergio Servantez}, \bibinfo{person}{Joe Barrow}, \bibinfo{person}{Kristian Hammond}, {and} \bibinfo{person}{Rajiv Jain}.} \bibinfo{year}{2024}\natexlab{}.
\newblock \showarticletitle{Chain of Logic: Rule-Based Reasoning with Large Language Models}.
\newblock \bibinfo{journal}{\emph{arXiv preprint arXiv:2402.10400}} (\bibinfo{year}{2024}).
\newblock


\bibitem[Spadini et~al\mbox{.}(2017)]%
        {spadini2017mock}
\bibfield{author}{\bibinfo{person}{Davide Spadini}, \bibinfo{person}{Maur{\'\i}cio Aniche}, \bibinfo{person}{Magiel Bruntink}, {and} \bibinfo{person}{Alberto Bacchelli}.} \bibinfo{year}{2017}\natexlab{}.
\newblock \showarticletitle{To mock or not to mock? an empirical study on mocking practices}. In \bibinfo{booktitle}{\emph{2017 IEEE/ACM 14th International Conference on Mining Software Repositories (MSR)}}. IEEE, \bibinfo{pages}{402--412}.
\newblock


\bibitem[Sun et~al\mbox{.}(2023)]%
        {sun2023evolutionary}
\bibfield{author}{\bibinfo{person}{Baicai Sun}, \bibinfo{person}{Dunwei Gong}, \bibinfo{person}{Feng Pan}, \bibinfo{person}{Xiangjuan Yao}, {and} \bibinfo{person}{Tian Tian}.} \bibinfo{year}{2023}\natexlab{}.
\newblock \showarticletitle{Evolutionary generation of test suites for multi-path coverage of MPI programs with non-determinism}.
\newblock \bibinfo{journal}{\emph{IEEE Transactions on Software Engineering}} \bibinfo{volume}{49}, \bibinfo{number}{6} (\bibinfo{year}{2023}), \bibinfo{pages}{3504--3523}.
\newblock


\bibitem[Tang et~al\mbox{.}(2024)]%
        {tang2024chatgpt}
\bibfield{author}{\bibinfo{person}{Yutian Tang}, \bibinfo{person}{Zhijie Liu}, \bibinfo{person}{Zhichao Zhou}, {and} \bibinfo{person}{Xiapu Luo}.} \bibinfo{year}{2024}\natexlab{}.
\newblock \showarticletitle{Chatgpt vs sbst: A comparative assessment of unit test suite generation}.
\newblock \bibinfo{journal}{\emph{IEEE Transactions on Software Engineering}} (\bibinfo{year}{2024}).
\newblock


\bibitem[Tonella(2004)]%
        {tonella2004evolutionary}
\bibfield{author}{\bibinfo{person}{Paolo Tonella}.} \bibinfo{year}{2004}\natexlab{}.
\newblock \showarticletitle{Evolutionary testing of classes}.
\newblock \bibinfo{journal}{\emph{ACM SIGSOFT Software Engineering Notes}} \bibinfo{volume}{29}, \bibinfo{number}{4} (\bibinfo{year}{2004}), \bibinfo{pages}{119--128}.
\newblock


\bibitem[Tufano et~al\mbox{.}(2020)]%
        {tufano2020unit}
\bibfield{author}{\bibinfo{person}{M. Tufano}, \bibinfo{person}{D. Drain}, \bibinfo{person}{A. Svyatkovskiy}, {and} \bibinfo{person}{et al.}} \bibinfo{year}{2020}\natexlab{}.
\newblock \showarticletitle{Unit test case generation with transformers and focal context}.
\newblock \bibinfo{journal}{\emph{arXiv preprint arXiv:2009.05617}} (\bibinfo{year}{2020}).
\newblock


\bibitem[Wang et~al\mbox{.}(2020)]%
        {wang2020automatic}
\bibfield{author}{\bibinfo{person}{Chunhui Wang}, \bibinfo{person}{Fabrizio Pastore}, \bibinfo{person}{Arda Goknil}, {and} \bibinfo{person}{Lionel~C Briand}.} \bibinfo{year}{2020}\natexlab{}.
\newblock \showarticletitle{Automatic generation of acceptance test cases from use case specifications: an nlp-based approach}.
\newblock \bibinfo{journal}{\emph{IEEE Transactions on Software Engineering}} \bibinfo{volume}{48}, \bibinfo{number}{2} (\bibinfo{year}{2020}), \bibinfo{pages}{585--616}.
\newblock


\bibitem[Wang et~al\mbox{.}(2024)]%
        {wang2024hits}
\bibfield{author}{\bibinfo{person}{Zejun Wang}, \bibinfo{person}{Kaibo Liu}, \bibinfo{person}{Ge Li}, {and} \bibinfo{person}{Zhi Jin}.} \bibinfo{year}{2024}\natexlab{}.
\newblock \bibinfo{title}{HITS: High-coverage LLM-based Unit Test Generation via Method Slicing}.
\newblock
\newblock


\bibitem[Wei et~al\mbox{.}(2022)]%
        {wei2022free}
\bibfield{author}{\bibinfo{person}{Anjiang Wei}, \bibinfo{person}{Yinlin Deng}, \bibinfo{person}{Chenyuan Yang}, {and} \bibinfo{person}{Lingming Zhang}.} \bibinfo{year}{2022}\natexlab{}.
\newblock \showarticletitle{Free lunch for testing: Fuzzing deep-learning libraries from open source}. In \bibinfo{booktitle}{\emph{Proceedings of the 44th International Conference on Software Engineering}}. \bibinfo{pages}{995--1007}.
\newblock


\bibitem[wikipedia(2024)]%
        {GWT}
\bibfield{author}{\bibinfo{person}{wikipedia}.} \bibinfo{year}{2024}\natexlab{}.
\newblock \bibinfo{title}{"Given-When-Then."}.
\newblock
\newblock
\urldef\tempurl%
\url{https://en.wikipedia.org/wiki/Given-When-Then}
\showURL{%
\tempurl}
\newblock
\shownote{Accessed: 2024}.


\bibitem[Xie et~al\mbox{.}(2023)]%
        {xie2023chatunitest}
\bibfield{author}{\bibinfo{person}{Zhuokui Xie}, \bibinfo{person}{Yinghao Chen}, \bibinfo{person}{Chen Zhi}, \bibinfo{person}{Shuiguang Deng}, {and} \bibinfo{person}{Jianwei Yin}.} \bibinfo{year}{2023}\natexlab{}.
\newblock \bibinfo{title}{ChatUniTest: a ChatGPT-based automated unit test generation tool}.
\newblock
\newblock


\bibitem[Xu et~al\mbox{.}(2024)]%
        {xu2024mr}
\bibfield{author}{\bibinfo{person}{Congying Xu}, \bibinfo{person}{Songqiang Chen}, \bibinfo{person}{Jiarong Wu}, \bibinfo{person}{Shing-Chi Cheung}, \bibinfo{person}{Valerio Terragni}, \bibinfo{person}{Hengcheng Zhu}, {and} \bibinfo{person}{Jialun Cao}.} \bibinfo{year}{2024}\natexlab{}.
\newblock \showarticletitle{MR-Adopt: Automatic Deduction of Input Transformation Function for Metamorphic Testing}.
\newblock \bibinfo{journal}{\emph{arXiv preprint arXiv:2408.15815}} (\bibinfo{year}{2024}).
\newblock


\bibitem[Yandrapally and Mesbah(2022)]%
        {yandrapally2022fragment}
\bibfield{author}{\bibinfo{person}{Rahul~Krishna Yandrapally} {and} \bibinfo{person}{Ali Mesbah}.} \bibinfo{year}{2022}\natexlab{}.
\newblock \showarticletitle{Fragment-based test generation for web apps}.
\newblock \bibinfo{journal}{\emph{IEEE Transactions on Software Engineering}} \bibinfo{volume}{49}, \bibinfo{number}{3} (\bibinfo{year}{2022}), \bibinfo{pages}{1086--1101}.
\newblock


\bibitem[Yang et~al\mbox{.}(2024)]%
        {yang2024enhancing}
\bibfield{author}{\bibinfo{person}{C. Yang}, \bibinfo{person}{J. Chen}, \bibinfo{person}{B. Lin}, {and} \bibinfo{person}{et al.}} \bibinfo{year}{2024}\natexlab{}.
\newblock \showarticletitle{Enhancing LLM-based Test Generation for Hard-to-Cover Branches via Program Analysis}.
\newblock \bibinfo{journal}{\emph{arXiv preprint arXiv:2404.04966}} (\bibinfo{year}{2024}).
\newblock


\bibitem[Ye et~al\mbox{.}(2023)]%
        {ye2023generative}
\bibfield{author}{\bibinfo{person}{Guixin Ye}, \bibinfo{person}{Tianmin Hu}, \bibinfo{person}{Zhanyong Tang}, \bibinfo{person}{Zhenye Fan}, \bibinfo{person}{Shin~Hwei Tan}, \bibinfo{person}{Bo Zhang}, \bibinfo{person}{Wenxiang Qian}, {and} \bibinfo{person}{Zheng Wang}.} \bibinfo{year}{2023}\natexlab{}.
\newblock \showarticletitle{A Generative and Mutational Approach for Synthesizing Bug-Exposing Test Cases to Guide Compiler Fuzzing}. In \bibinfo{booktitle}{\emph{Proceedings of the 31st ACM Joint European Software Engineering Conference and Symposium on the Foundations of Software Engineering}}. \bibinfo{pages}{1127--1139}.
\newblock


\bibitem[Yu et~al\mbox{.}(2024)]%
        {yu2024practitioners}
\bibfield{author}{\bibinfo{person}{Xiao Yu}, \bibinfo{person}{Lei Liu}, \bibinfo{person}{Xing Hu}, \bibinfo{person}{Jacky Keung}, \bibinfo{person}{Xin Xia}, {and} \bibinfo{person}{David Lo}.} \bibinfo{year}{2024}\natexlab{}.
\newblock \showarticletitle{Practitioners’ Expectations on Automated Test Generation}. In \bibinfo{booktitle}{\emph{Proceedings of the 33rd ACM SIGSOFT International Symposium on Software Testing and Analysis}}. \bibinfo{pages}{1618--1630}.
\newblock


\bibitem[Yuan et~al\mbox{.}(2023)]%
        {yuan2023no}
\bibfield{author}{\bibinfo{person}{Zhiqiang Yuan}, \bibinfo{person}{Yiling Lou}, \bibinfo{person}{Mingwei Liu}, \bibinfo{person}{Shiji Ding}, \bibinfo{person}{Kaixin Wang}, \bibinfo{person}{Yixuan Chen}, {and} \bibinfo{person}{Xin Peng}.} \bibinfo{year}{2023}\natexlab{}.
\newblock \bibinfo{title}{No more manual tests? evaluating and improving chatgpt for unit test generation}.
\newblock
\newblock


\bibitem[Zhang et~al\mbox{.}(2023)]%
        {zhang2023repocoder}
\bibfield{author}{\bibinfo{person}{Fengji Zhang}, \bibinfo{person}{Bei Chen}, \bibinfo{person}{Yue Zhang}, \bibinfo{person}{Jacky Keung}, \bibinfo{person}{Jin Liu}, \bibinfo{person}{Daoguang Zan}, \bibinfo{person}{Yi Mao}, \bibinfo{person}{Jian-Guang Lou}, {and} \bibinfo{person}{Weizhu Chen}.} \bibinfo{year}{2023}\natexlab{}.
\newblock \showarticletitle{Repocoder: Repository-level code completion through iterative retrieval and generation}.
\newblock \bibinfo{journal}{\emph{arXiv preprint arXiv:2303.12570}} (\bibinfo{year}{2023}).
\newblock


\bibitem[Zhang et~al\mbox{.}(2024)]%
        {zhang2024autocoderover}
\bibfield{author}{\bibinfo{person}{Yuntong Zhang}, \bibinfo{person}{Haifeng Ruan}, \bibinfo{person}{Zhiyu Fan}, {and} \bibinfo{person}{Abhik Roychoudhury}.} \bibinfo{year}{2024}\natexlab{}.
\newblock \showarticletitle{Autocoderover: Autonomous program improvement}. In \bibinfo{booktitle}{\emph{Proceedings of the 33rd ACM SIGSOFT International Symposium on Software Testing and Analysis}}. \bibinfo{pages}{1592--1604}.
\newblock


\bibitem[Zhao et~al\mbox{.}(2022)]%
        {zhao2022avgust}
\bibfield{author}{\bibinfo{person}{Yixue Zhao}, \bibinfo{person}{Saghar Talebipour}, \bibinfo{person}{Kesina Baral}, \bibinfo{person}{Hyojae Park}, \bibinfo{person}{Leon Yee}, \bibinfo{person}{Safwat~Ali Khan}, \bibinfo{person}{Yuriy Brun}, \bibinfo{person}{Nenad Medvidovi{\'c}}, {and} \bibinfo{person}{Kevin Moran}.} \bibinfo{year}{2022}\natexlab{}.
\newblock \showarticletitle{Avgust: automating usage-based test generation from videos of app executions}. In \bibinfo{booktitle}{\emph{Proceedings of the 30th ACM Joint European Software Engineering Conference and Symposium on the Foundations of Software Engineering}}. \bibinfo{pages}{421--433}.
\newblock


\bibitem[Zhou et~al\mbox{.}(2022)]%
        {zhou2022selectively}
\bibfield{author}{\bibinfo{person}{Zhichao Zhou}, \bibinfo{person}{Yuming Zhou}, \bibinfo{person}{Chunrong Fang}, \bibinfo{person}{Zhenyu Chen}, {and} \bibinfo{person}{Yutian Tang}.} \bibinfo{year}{2022}\natexlab{}.
\newblock \showarticletitle{Selectively combining multiple coverage goals in search-based unit test generation}. In \bibinfo{booktitle}{\emph{Proceedings of the 37th IEEE/ACM International Conference on Automated Software Engineering}}. \bibinfo{pages}{1--12}.
\newblock


\bibitem[Zhu et~al\mbox{.}(2024)]%
        {zhu2024deepseek}
\bibfield{author}{\bibinfo{person}{Qihao Zhu}, \bibinfo{person}{Daya Guo}, \bibinfo{person}{Zhihong Shao}, \bibinfo{person}{Dejian Yang}, \bibinfo{person}{Peiyi Wang}, \bibinfo{person}{Runxin Xu}, \bibinfo{person}{Y Wu}, \bibinfo{person}{Yukun Li}, \bibinfo{person}{Huazuo Gao}, \bibinfo{person}{Shirong Ma}, {et~al\mbox{.}}} \bibinfo{year}{2024}\natexlab{}.
\newblock \bibinfo{title}{DeepSeek-Coder-V2: Breaking the Barrier of Closed-Source Models in Code Intelligence}.
\newblock
\newblock


\end{thebibliography}

\appendix

\end{document}